\documentclass[11pt, oneside]{article} 
\pdfoutput=1  	

\usepackage{jheppub}

\usepackage{amsmath}
\usepackage{amssymb}
\usepackage{amsfonts}
\usepackage{amsthm}
\usepackage{amsbsy}
\usepackage{array}
\usepackage[linesnumbered,lined,boxed,commentsnumbered]{algorithm2e}

\newenvironment{lyxcode}
	{\par\begin{list}{}{
		\setlength{\rightmargin}{\leftmargin}
		\setlength{\listparindent}{0pt}
		\raggedright
		\setlength{\itemsep}{0pt}
		\setlength{\parsep}{0pt}
		\normalfont\ttfamily}%
	 \item[]}
	{\end{list}}

\usepackage[OT2,OT1]{fontenc}
\newcommand\cyr
{
\renewcommand\rmdefault{wncyr}
\renewcommand\sfdefault{wncyss}
\renewcommand\encodingdefault{OT2}
\normalfont
\selectfont
}

\DeclareTextFontCommand{\textcyr}{\cyr}

\usepackage{mathtools}
\usepackage[usenames,dvipsnames]{xcolor}

\usepackage{subcaption}

\usepackage{dsdshorthand}

\usepackage[vcentermath]{youngtab}

\makeatletter
\def\@fpheader{\ }
\makeatother

\title{\texttt{blocks\_3d}: Software for general 3d conformal blocks} 

\author{ Rajeev S. Erramilli$^a$, Luca V. Iliesiu$^{b,c}$, Petr Kravchuk$^d$, Walter Landry$^{e,f}$, David Poland$^{a}$,
David Simmons-Duffin$^{e}$\\}

\affiliation{${}^a$Department of Physics, Yale University, New Haven, CT 06520, USA \\
${}^b$Joseph Henry Laboratories, Princeton University, Princeton, NJ 08544, USA \\
${}^c$Stanford Institute for Theoretical Physics, Stanford University, Stanford, CA 94305 \\
${}^d$School of Natural Sciences, Institute for Advanced Study, Princeton, NJ 08540, USA\\
${}^e$Walter Burke Institute for Theoretical Physics, Caltech, Pasadena, CA 91125, USA\\
${}^f$Simons Collaboration on the Nonperturbative Bootstrap\\
}

\abstract{ We introduce the software \texttt{blocks\_3d} for computing
  four-point conformal blocks of operators with arbitrary Lorentz
  representations in 3d CFTs. It uses Zamolodchikov-like recursion
  relations to numerically compute derivatives of blocks around a
  crossing-symmetric configuration. It is implemented as a heavily
  optimized, multithreaded, \texttt{C++} application.  We give
  performance benchmarks for correlators containing scalars, fermions,
  and stress tensors. As an example application, we recompute
  bootstrap bounds on four-point functions of fermions and study whether a previously observed sharp jump can be explained using the
  ``fake primary" effect. We conclude that the fake primary effect cannot fully explain the jump and the possible existence of a ``dead-end" CFT near the jump merits further study.  } \date{}
\preprint{}

\preprint{CALT-TH 2020-048}

\begin{document}

\maketitle

\section{Introduction}

The conformal bootstrap has emerged as a powerful tool for the nonperturbative study of conformal field theories (CFTs). In particular, numerical studies of bootstrap equations~\cite{Rattazzi:2008pe} using semidefinite programming methods~\cite{Poland:2011ey,Simmons-Duffin:2015qma,Landry:2019qug} have led to precise determinations of CFT data in a variety of interesting theories, including the 3d Ising~\cite{ElShowk:2012ht, El-Showk:2014dwa, Kos:2014bka, Kos:2016ysd, Simmons-Duffin:2016wlq} and $\mathrm{O}(N)$~\cite{Kos:2013tga, Kos:2015mba, Kos:2016ysd, Chester:2019ifh,Liu:2020tpf} models. For a review of recent developments, see~\cite{Poland:2018epd}. 

Most studies pursued so far in $d>2$ have focused on correlation functions of scalar operators, where the conformal blocks appearing in the bootstrap equations are relatively easy to compute. In 4d CFTs, they are expressible in terms of hypergeometric functions~\cite{Dolan:2000ut}, while in 3d CFTs, they can be straightforwardly computed using Zamolodchikov-like recursion relations developed in~\cite{Kos:2013tga,Kos:2014bka,Penedones:2015aga,Yamazaki:2016vqi}. These recursion relations have been implemented efficiently in the software package \texttt{scalar\_blocks}~\cite{ScalarBlocks}.

A handful of studies have been carried out for correlation functions of spinning operators, including four-point functions of fermions~\cite{Iliesiu:2015qra,Iliesiu:2017nrv,Karateev:2019pvw}, stress-tensors~\cite{Dymarsky:2017yzx}, currents~\cite{Dymarsky:2017xzb}, and mixed correlators containing scalars and currents~\cite{Reehorst:2019pzi}. Each of these studies faced a huge technical hurdle of understanding how to compute the relevant conformal blocks for each combination of three-point and four-point tensor structures contributing to the correlator and then carrying out the computation in practice. These studies solved these problems on an ad hoc basis, employing a variety of different methods that do not easily scale to larger problems. 

Recently, a general recursive algorithm for computing 3d conformal blocks of arbitrary spin was introduced in~\cite{Erramilli:2019njx}, building on the earlier recursion relations~\cite{Penedones:2015aga, Kravchuk:2017dzd}. The algorithm uses the fact that the conformal blocks have an expansion in poles in the exchanged scaling dimension:
\be
	g^{ab}_{\De,j,I}(z,\bar z)\sim \frac{1}{\De-\De_{j,i}}(\cL_{j,i})^a_{a'}(\cR_{j,i})^b_{b'}g^{a'b'}_{\De'_{j,i},j'_{j,i},I}(z,\bar z),
\ee
where $\Delta_{j,i}$ describe a known infinite set of poles and the residues are themselves conformal blocks up to some additional residue matrices $\cL_{j,i}, \cR_{j,i}$. In addition to the scaling dimension $\Delta$ and spin $j$, each block is labeled by a pair of three-point structures $ab$ and a four-point structure $I$. General formulas for these matrices, as well as formulas for the $\Delta \rightarrow \infty$ limit of the blocks needed to implement the recursion relation, were computed explicitly in~\cite{Erramilli:2019njx}. 

A variety of different bases have been employed to describe the three-and four-point structures, including polynomial or differential bases in embedding-space structures~\cite{Costa:2011mg,Costa:2011dw,SimmonsDuffin:2012uy,Costa:2014rya,Elkhidir:2014woa,Iliesiu:2015qra,Echeverri:2015rwa,Costa:2016hju,Karateev:2017jgd,Fortin:2019dnq,Fortin:2019pep,Fortin:2019gck,Fortin:2020ncr}, and a $q$-basis which is naturally defined using the conformal frame approach~\cite{Kravchuk:2016qvl}. In~\cite{Erramilli:2019njx} an additional basis was introduced, called the SO(3) basis, where the matrices $\cL_{j,i}, \cR_{j,i}$ take on a particularly simple block-diagonal form. For three-point functions between operators of SO(3) spins $j_1, j_2, j_3$, this basis can be labeled by the possible spins $(j_{12}, j_{123})$ appearing in the decompositions $j_{12} \in j_1 \otimes j_2$ and $j_{123} \in j_{12} \otimes j_3$. 

In this work, we present the software package \texttt{blocks\_3d}, which efficiently implements the recursive algorithm of~\cite{Erramilli:2019njx}. In particular, given external operators of spins $j_1, j_2, j_3, j_4$, a four-point structure specified in the $q$-basis, and the spin combinations $j_{12}$ and $j_{43}$, the software computes conformal block derivatives (using arbitrary-precision arithmetic) up to a specified recursion order for all allowed SO(3)-basis structures. Derivatives of conformal blocks around the crossing-symmetric configuration are returned up to a cutoff $\Lambda$, where derivatives can be taken in several different coordinates. We give performance benchmarks for the code, comparing our code first to \texttt{scalar\_blocks}.  We then test it on several correlators containing scalars, fermions, and stress tensors, demonstrating that \texttt{blocks\_3d} is feasible to use in large-scale numerical bootstrap calculations with spinning correlators. 

We further explicitly demonstrate how to use the code in the context of the bootstrap for four-point functions of 3d Majorana fermions $\<\psi\psi\psi\psi\>$.  We reproduce bounds on the leading parity-even operator $\epsilon$ and parity-odd scalar $\sigma$ appearing in the $\psi \times \psi$ operator product expansion, previously obtained in~\cite{Iliesiu:2015qra}. The former shows a prominent kink, and the latter shows a sharp jump at the same value of $\Delta_{\psi}$. This jump was previously conjectured to relate to the ``fake primary" effect described in~\cite{Karateev:2019pvw}, where a parity-odd spin-1 operator $V$ at the unitarity bound can mimic a scalar of dimension $\Delta_{\sigma} = 3$. We impose a small gap in the spin-1 sector and observe that the jump and kink are stable, persisting up to $\Delta_{V} \sim 2.3$. Our tentative conclusion is that the feature is unlikely to be fully explained by the fake primary effect and merits further study.

This paper is organized as follows. In section~\ref{sec:preliminaries} we describe our conventions and the bases of tensor structures that we use. We also give an exact formula for the conformal block corresponding to identity exchange in these conventions. In section~\ref{sec:blocks3d} we describe the algorithm used by \texttt{blocks\_3d}, the structure of the output, and the details of the implementation. In section~\ref{sec:performance} we give performance benchmarks, and in section~\ref{sec:example} we describe in detail an example of using the software for the 4-fermion bootstrap. We conclude in section~\ref{sec:conclusions} by describing some possible future applications of \texttt{blocks\_3d}. Appendices contain the link to \texttt{blocks\_3d} code as well as the conventions and details omitted in the main text.

\section{Mathematical preliminaries}
\label{sec:preliminaries}

\subsection{Definition of conformal blocks}
\label{sec:defn}

In this section we give a precise definition of the conformal blocks that are computed by \texttt{blocks\_3d}. Since we will only be interested in conformal blocks in 3d CFTs, we will not be any more general than required. We will work in Lorentzian signature since unitarity is the most manifest there.

Consider the Hermitian local primary operators $\cO_i$ ($i=1,\cdots,4$) and $\cO$. Let their scaling dimensions and spins be $(\De_i,j_i)$ and $(\De,j)$, respectively. The spins $j,j_i$ can be integer or half-integer; we do \textit{not} restrict to bosonic representations.\footnote{In particular, the ordering of operators in the definitions plays an important role due to possible signs from permutations of fermions.} We work with the following realization of the spin-$j$ representations: the operator $\cO$ carries $2j$ spinor indices $\a_k$,
\be
	\cO^{\a_1\cdots\a_{2j}}(x),
\ee
and is completely symmetric in these indices. We choose our conventions (see details in appendix~\ref{app:spinorconv}) so that the representation matrices of the Lorentz group~$\mathrm{Spin}(2,1)$ are real when acting on $\a_k$, and thus we can indeed assume that $\cO$ is Hermitian,
\be
\label{eq:hermitian-prop-cO}
	(\cO^{\a_1\cdots\a_{2j}}(x))^\dagger=	
	\cO^{\a_1\cdots\a_{2j}}(x).
\ee
The same comments apply to $\cO_i$. Furthermore, we assume that $\cO$ is normalized so that its time-ordered two-point function for space-like separated $x_1, x_2$ is given by
\be\label{eq:twoptnormalization}
	\<\cO(x_1,s_1)\cO(x_2,s_2)\>=c_\cO\frac{i^{2j}(s_1^\a\g^\mu_{\a\b}s_2^\b x_{12,\mu})^{2j}}{x_{12}^{2\De+2j}},
\ee
where $x_{12}=x_1-x_2$, $c_\cO$ is a positive constant defined below, and we used the index-free notation
\be
	\cO(x,s)=\cO^{\a_1\cdots\a_{2j}}(x)s_{\a_1}\cdots s_{\a_{2j}}
\ee
for an auxiliary spinor $s$. The constant $c_\cO$ is given by
\be\label{eq:2ptnorm}
	c_\cO=(4/a_0)^\De b_j,
\ee
where we leave the choice of $a_0$, and the spin-dependent $b_j>0$, up to the user. Note that unlike $c_\cO$, the phase in~\eqref{eq:twoptnormalization} is fixed
by unitarity, i.e. requiring that the two-point function defines a positive-definite norm on the Hilbert space. The unitarity bounds on $\De$ are
\be
	\begin{cases}
		\De=0\text{ or }\De\geq\half, & j=0, \\
		\De\geq 1, & j=\half, \\
		\De\geq j+1, & j > \half.
	\end{cases}
\ee
When $\De$ is in the interior of these bounds, i.e.
\be
	\begin{cases}
		\De>\half, & j=0, \\
		\De>1, & j=\half, \\
		\De>j+1, & j > \half,
	\end{cases}
\ee
the conformal block is guaranteed to be finite.

We parametrize the values of the three-point functions of the operators $\cO_i,\cO$ in the following way,
\be
	\<\cO_1(x_1,s_1)\cO_2(x_2,s_2)\cO(x_3,s_3)\>&=\sum_{a\in \cI_{12\cO}} \l_{12\cO,(a)}\<\cO_1(x_1,s_1)\cO_2(x_2,s_2)\cO(x_3,s_3)\>^{(a)},\nn\\
	\<\cO_4(x_1,s_1)\cO_3(x_2,s_2)\cO(x_3,s_3)\>&=\sum_{a\in \cI_{43\cO}} \l_{43\cO,(a)}\<\cO_4(x_1,s_1)\cO_3(x_2,s_2)\cO(x_3,s_3)\>^{(a)},
\ee
where our choice of standard conformally invariant three-point tensor structures 
\be
\<\cO_1(x_1,s_1)\cO_2(x_2,s_2)\cO(x_3,s_3)\>^{(a)},
\<\cO_4(x_1,s_1)\cO_3(x_2,s_2)\cO(x_3,s_3)\>^{(a)},
\ee
which are labeled by indices in some sets $\cI_{12\cO}$ and $\cI_{43\cO}$, is described in section~\ref{sec:3ptbasis}. Similarly, the four-point function is decomposed as
\be
	&\<\cO_1(x_1,s_1)\cO_2(x_2,s_2)\cO_3(x_3,s_3)\cO_4(x_4,s_4)\>\nn\\
	&=\sum_{I\in\cI_{1234}}\<\cO_1(x_1,s_1)\cO_2(x_2,s_2)\cO_3(x_3,s_3)\cO_4(x_4,s_4)\>^{(I)}g_I(z,\bar z),
\ee
where our choice of standard conformally invariant four-point tensor structures,
\be
\<\cO_1(x_1,s_1)\cO_2(x_2,s_2)\cO_3(x_3,s_3)\cO_4(x_4,s_4)\>^{(I)},
\ee
and the cross-ratios $z,\bar z$, are defined in section~\ref{sec:4ptstructs}, and the index $I$ runs over some index set $\cI_{1234}$.

With this introduction, the conformal block $g^{ab}_{\De,j,I}(z,\bar z)$ is defined by requiring that the contribution of $\cO$ and its descendants to the $\cO_1\times \cO_2$ (equivalently, $\cO_3\times \cO_4$) OPE in the above four-point function is given by
\be
	g_I(z,\bar z)=\cdots + \sum_{a\in \cI_{12\cO}} \,\sum_{b\in \cI_{43\cO}}  \l_{12\cO,(a)}\l_{43\cO,(b)}g^{ab}_{\De,j,I}(z,\bar z)+\cdots,
\ee
where $\cdots$ denote contributions from other primary operators and descendants. When we use the notation $g^{ab}_{\De,j,I}(z,\bar z)$ there is implicit dependence on the scaling dimensions and spins $\De_i,j_i$. 

This definition, as well as the choices of three- and four-point structure bases described in sections~\ref{sec:3ptbasis} and~\ref{sec:4ptstructs}, are the same as in~\cite{Erramilli:2019njx}.

\subsection{Three-point structure basis}
\label{sec:3ptbasis}

In this section we define two bases of three-point structures, the $q$-basis (first introduced in~\cite{Kravchuk:2016qvl}) and the $\SO(3)$-basis (first introduced in~\cite{Karateev:2018oml,Erramilli:2019njx}). We introduce two bases because the physical properties of the three-point structures are easier to understand in the $q$-basis, while \texttt{blocks\_3d}, for performance reasons explained below, uses the $\SO(3)$-basis.

Let us explain in more detail what we mean by ``defining a basis of structures.'' Defining a basis of three-point structures means to provide an algorithm which, given some quantum numbers $\De_i,j_i$, produces a finite set $\cI_{123}$ of functions
\be
	f^{(a)}(x_1,s_1;x_2,s_2;x_3,s_3)
\ee
which are linearly-independent, invariant under conformal transformations, and form a complete basis for functions of this form. Here, it is understood that conformal transformations act on $f^{(a)}$ in the same way as on
\be
	\<\cO_1(x_1,s_1)\cO_2(x_2,s_2)\cO_3(x_3,s_3)\>,
\ee
where primary operators $\cO_i$ have scaling dimension and spin $\De_i,j_i$. We have already used, and will be using in what follows, a somewhat misleading notation for $f^{(a)}$,
\be\label{eq:structnotation}
\<\cO_1(x_1,s_1)\cO_2(x_2,s_2)\cO_3(x_3,s_3)\>^{(a)}\equiv f^{(a)}(x_1,s_1;x_2,s_2;x_3,s_3).
\ee
The reason why this notation is misleading is that the left-hand side appears to depend on the concrete primary operators $\cO_i$ and suggests that their ordering is somehow related to the ordering of operators in physical correlators. Instead, $f^{(a)}$ only depends on the quantum numbers $\De_i,j_i$. While the ordering of these quantum numbers is important, it has nothing to do with the order of operators in correlation functions. Nevertheless, the above notation allows us to quickly summarize the quantum numbers that the tensor structures correspond to, and for this reason we prefer to use it. We hope this won't cause too much confusion.

\subsubsection{$q$-basis}

In this subsection we summarize the definition and properties of the $q$-basis of three-point structures. We don't give any proofs, for which we instead refer the reader to~\cite{Kravchuk:2016qvl}. 

The $q$-basis structures for quantum numbers $(\De_i,j_i)$ are labeled by triples $a=[q_1,q_2,q_3]$, where $q_i$ are (half-)integers which range over
\be
	q_i\in \{-j_i,-j_i+1,\cdots, j_i\},
\ee
and are subject to
\be
	q_1+q_2+q_3=0.
\ee
The tensor structures 
\be
	\<\cO_1(x_1,s_1)\cO_2(x_2,s_2)\cO_3(x_3,s_3)\>^{[q_1q_2q_3]}
\ee
are fixed uniquely by conformal invariance and the requirement that for
\be\label{eq:3ptframe}
	x_1&=(0,0,0),\\
	x_2&=(0,0,1),\\
	x_3(L)&=(0,0,L),
\ee
the following identity holds
\be\label{eq:qdefn}
	&\lim_{L\to +\oo} L^{2\De_3}\<\cO_1(x_1,s_1)\cO_2(x_2,s_2)\cO_3(x_3(L),s_3)\>^{[q_1q_2q_3]}
	=\prod_{i=1}^3 ((s_i)_1)^{j_i+q_i}((s_i)_2)^{j_i-q_i}.
\ee

This definition is rather implicit. However, it makes it easy to convert between the $q$-basis and more explicit structures such as those introduced in~\cite{Costa:2011mg,Iliesiu:2015qra}: we just have to evaluate these explicit tensor structures in the same way as in the left hand side of~\eqref{eq:qdefn}, and express the result as a linear combination of the monomials appearing in the right-hand side of~\eqref{eq:qdefn}. Notice that this definition is not permutation-invariant: the representations $(\De_i,j_i)$ that the operators $\cO_i$ symbolize are listed in a particular order, and the respective coordinates are set to different values above. For example, it is the coordinate corresponding to the last representation that is sent to infinity. We stress that the order is determined by the order in which the representations (represented by $\cO_i$) are listed, and not by their indices $1,2,3$. This is important for understanding the meaning of permutation properties below. 

Let us list some simple properties of these structures~\cite{Kravchuk:2016qvl}:
\begin{itemize}
	\item If we expand a three-point function of Hermitian operators in the $q$-basis, then the OPE coefficients are real if all $j_i$ are integers and pure imaginary otherwise.
	\item Space parity acts on the $q$-basis structures as 
	\be
	\label{eq:parity-on-q-struct}
		&\<\cO_1(x_1,s_1)\cO_2(x_2,s_2)\cO_3(x_3,s_3)\>^{[q_1q_2q_3]}\nn\\
		&\to \<\cO_1(x_1,s_1)\cO_2(x_2,s_2)\cO_3(x_3,s_3)\>^{[-q_1,-q_2,-q_3]}.
	\ee
	See appendix~\ref{app:parity-conv} for details on what is meant by ``space parity''.
	\item $q$-basis structures have the following permutation properties under transpositions
	\be
	&\<\cO_1(x_1,s_1)\cO_2(x_2,s_2)\cO_3(x_3,s_3)\>^{[q_1q_2q_3]}\nn\\
	&=(-1)^{j_1+j_2-j_3}\<\cO_2(x_2,s_2)\cO_1(x_1,s_1)\cO_3(x_3,s_3)\>^{[-q_2,-q_1,-q_3]},\\
	&=(-1)^{j_1+j_2+j_3}\<\cO_3(x_3,s_3)\cO_2(x_2,s_2)\cO_1(x_1,s_1)\>^{[-q_3,-q_2,-q_1]},\\
	&=(-1)^{-j_1+j_2+j_3}\<\cO_1(x_1,s_1)\cO_3(x_3,s_3)\cO_2(x_2,s_2)\>^{[-q_1,-q_3,-q_2]}.
	\ee
	Other permutations can be obtained by composing these basic transpositions. To understand the precise meaning of these equations, recall the discussion around~\eqref{eq:structnotation}. For example, the first equation says that if we take the basis structure labeled by $[q_1,q_2,q_3]$, constructed for representations $(\De_1,j_1),(\De_2,j_2),(\De_3,j_3)$, and evaluate it at $(x_1,s_1), (x_2,s_2), (x_3,s_3)$, then it is equal to $(-1)^{j_1+j_2-j_3}$ times the basis structure labeled by $[-q_2,-q_1,-q_3]$, constructed for representations $(\De_2,j_2),(\De_1,j_1),(\De_3,j_3)$, and evaluated at $(x_2,s_2), (x_1,s_1), (x_3,s_3)$.
\end{itemize}
Because of the way permutations and parity act on these structures, we will often consider the structures defined as
\be
&\<\cO_1(x_1,s_1)\cO_2(x_2,s_2)\cO_3(x_3,s_3)\>^{[q_1q_2q_3]^\pm}\nn\\
&\equiv
\<\cO_1(x_1,s_1)\cO_2(x_2,s_2)\cO_3(x_3,s_3)\>^{[q_1q_2q_3]}\pm \<\cO_1(x_1,s_1)\cO_2(x_2,s_2)\cO_3(x_3,s_3)\>^{[-q_1,-q_2,-q_3]}.
\ee
The structures with $(+)$ sign are parity-even and those with $(-)$ sign are parity-odd. Furthermore, transpositions simply permute the $q_i$ in the labels of these structures, but for the $(-)$ structures we get an extra factor of $(-1)$ in the action of transpositions.

We illustrate how these structures can be used in practice in the example of the four-fermion bootstrap in section~\ref{sec:example}.

\subsubsection{$\SO(3)$-basis}
\label{sec:SO3-basis}

In this subsection we define the $\SO(3)$ basis of structures, first introduced in~\cite{Karateev:2018oml,Erramilli:2019njx}.\footnote{In~\cite{Erramilli:2019njx} this basis was called the $\SO(3)_r$ basis, and the $\SO(3)$ basis was referred to as an intermediate basis. Since in this paper, we simplify the construction of~\cite{Erramilli:2019njx}, we will, for simplicity, call the final basis the $\SO(3)$ basis. We hope this won't cause confusion.} We will only give the formal expressions for the structures and not explain the motivation behind them, for which we refer the reader to~\cite{Erramilli:2019njx}.

First, we define the following monomials in $s_i$
\be\label{eq:so2defn}
	|j_1,m_1;j_2,m_2;j_3,m_3\>\equiv (-1)^{j_1-j_3+m_2} \prod_{i=1}^3 \binom{2j_i}{j_i+m_i}^{1/2}((s_i)_1)^{j_i+m_i}((s_i)_2)^{j_i-m_i},
\ee
where $m_i$ can take the same values as $q_i$,
\be
	&m_i\in\{-j_i,-j_i+1,\cdots,j_i\},\\
	&m_1+m_2+m_3=0.
\ee
Note that these monomials are proportional to the ones appearing in the right-hand side of~\eqref{eq:qdefn} with $m_i=q_i$. As the notation suggests, these monomials transform in the standard way under some $\mathfrak{su}(2)$ algebra. With this in mind, we define
\be\label{eq:so2.5defn}
	|j_{12},-m_3;j_3,m_3\>&\equiv \sum_{m_1,m_2\atop m_1+m_2=-m_3}\<j_1,m_1;j_2,m_2|j_{12},-m_3\>|j_1,m_1;j_2,m_2;j_3,m_3\>,\\
	|j_{12},j_{123}\>&\equiv \sum_{m_3} \<j_{12},-m_3;j,m_3|j_{123},0\>|j_{12},-m_3;j_3,m_3\>,\label{eq:so3polydefn}
\ee
where $\<j_1,m_1;j_2,m_2|j,m\>$ are the Clebsch-Gordan coefficients. The inverse formulas are
\be
\label{eq:conversionSO3toq-1}
	|j_{12},-m_3;j_3,m_3\>&=\sum_{j_{123}} \<j_{123},0|j_{12},-m_3;j,m_3\>|j_{12},j_{123}\>,\\
	\label{eq:conversionSO3toq-2}
	|j_1,m_1;j_2,m_2;j_3,m_3\>&=\sum_{j_{12}}\<j_{12},-m_3|j_1,m_1;j_2,m_2\>|j_{12},-m_3;j_3,m_3\>.
\ee
In all of the above formulas the summation is performed over the values of the variables for which the Clebsch-Gordan coefficients are non-vanishing. We use conventions for the Clebsch-Gordan coefficients such that they are real, so that
\be
	\<j_1,m_1;j_2,m_2|j,m\> = \<j,m|j_1,m_1;j_2,m_2\>.
\ee
To completely specify the conventions, we give a formula for Clebsch-Gordan coefficients in appendix~\ref{app:CG-coeff}.

Note that according to the above definitions, $|j_{12},j_{123}\>$ are simply polynomials in $s_i$. We define $\SO(3)$-basis structures analogously to~\eqref{eq:qdefn},
\be\label{eq:SO3defn}
	&\lim_{L\to +\oo} L^{2\De_3}\<\cO_1(x_1,s_1)\cO_2(x_2,s_2)\cO_3(x_3(L),s_3)\>^{(j_{12},j_{123})}
	=|j_{12},j_{123}\>,
\ee
where the values of $x_i$ are as in~\eqref{eq:3ptframe}. That is, the $\SO(3)$ basis structures are labeled by pairs $(j_{12},j_{123})$ for which the polynomials $|j_{12},j_{123}\>$ are non-zero, i.e.
\be
	j_{12}&\in \{|j_1-j_2|,|j_1-j_2|+1,\cdots j_1+j_2\},\\
	j_{123}&\in \{|j_{3}-j_{12}|,|j_{3}-j_{12}|+1,\cdots j_3+j_{12}\}.
\ee

By examining the definition of polynomials $|j_{12},j_{123}\>$, we see that they are explicitly defined as linear combinations of the monomials in the right-hand side of~\eqref{eq:qdefn} with coefficients given by products of Clebsch-Gordan coefficients. This definition implies that we can directly write the $\SO(3)$-basis structures as linear combinations of $q$-basis structures and vice versa. Since these expressions are rather bulky yet straightforward combinations of~\eqref{eq:so2defn},~\eqref{eq:so2.5defn} and~\eqref{eq:so3polydefn}, we omit them.

Since the Clebsch-Gordan coefficients are real, the reality properties of the OPE coefficients in $\SO(3)$ basis are the same as in $q$-basis. Most permutation properties are unfortunately not manifest in $\SO(3)$ basis. The parity of $\SO(3)$ structures is, on the other hand, manifest and is given by the sign of
\be
(-1)^{j_1-j_2+j_3-j_{123}}.
\ee

\subsection{Four-point structure basis}
\label{sec:4ptstructs}

In this section we define the $q$-basis for four-point tensor structures~\cite{Kravchuk:2016qvl}. The same comments as in section~\ref{sec:3ptbasis} apply to the meaning of the notation that we use for four-point tensor structures,
\be
	\<\cO_1(x_1,s_1)\cO_2(x_2,s_2)\cO_3(x_3,s_3)\cO_4(x_4,s_4)\>^{(I)}.
\ee
The $q$-basis structures are labeled by indices $I=[q_1,q_2,q_3,q_4]$ subject to\footnote{Note that unlike in the case of three-point structures there is no condition on $\sum_{i=1}^4q_i$.}
\be
	q_i\in\{-j_i,-j_i+1,\cdots, j_i\}.
\ee
A $q$-basis four-point tensor structure is defined by conformal invariance and its value in a standard configuration. Specifically, let
\be\label{eq:4ptCF}
	x_1&=(0,0,0),\\
	x_2&=(\tfrac{\bar z-z}{2},\tfrac{\bar z+z}{2},0),\\
	x_3&=(0,1,0),\\
	x_4(L)&=(0,L,0),
\ee
then we require
\be\label{eq:4ptqdefn}
	&\lim_{L\to +\oo}L^{2\De_4}\<\cO_1(x_1,s_1)\cO_2(x_2,s_2)\cO_3(x_3,s_3)\cO_4(x_4(L),s_4)\>^{[q_1q_2q_3q_4]}\nn\\
	&=\prod_{i=1}^4 ((s_i)_1)^{j_i+q_i}((s_i)_2)^{j_i-q_i}.
\ee
In particular, the decomposition of a four-point function into these structures can be computed by evaluating it in the configuration~\eqref{eq:4ptCF} and taking the limit as in the left-hand side of~\eqref{eq:4ptqdefn}. For the purposes of numerical bootstrap, it suffices to assume that $0<z,\bar z<1$, and so all operators are spacelike-separated. That is, the precise definition of $g_{[q_1q_2q_3q_4]}(z,\bar z)$ for $0<z,\bar z<1$ is 
\be
	&\lim_{L\to +\oo}L^{2\De_4}\<\cO_1(x_1,s_1)\cO_2(x_2,s_2)\cO_3(x_3,s_3)\cO_4(x_4(L),s_4)\>\nn\\
	&=\sum_{q_i} g_{[q_1q_2q_3q_4]}(z,\bar z)\prod_{i=1}^4 ((s_i)_1)^{j_i+q_i}((s_i)_2)^{j_i-q_i}.
\ee

Let us list some simple properties of the $q$-basis four-point tensor structures~\cite{Kravchuk:2016qvl}:
\begin{itemize}
	\item If we expand a four-point function of Hermitian operators in $q$-basis, then the coefficient functions are real if there are 0 or 4 fermions in the correlator, and imaginary if there are 2 fermions.
	\item The space parity of the $q$-basis stuctures is equal to $(-1)^{\sum_{i}j_i-q_i}$.
	See appendix~\ref{app:parity-conv} for details on what is meant by ``space parity''.
	\item Four-point $q$-basis tensor structures have simple properties under permutations for which we refer the reader to~\cite{Kravchuk:2016qvl}. The same comments about the meaning of permutations apply as in section~\ref{sec:3ptbasis}.
	\item The coefficient functions $g_{[q_1q_2q_3q_4]}(z,\bar z)$ transform in the following way under $z\leftrightarrow \bar z$,
	\be
	\label{eq:zzbsym}
	g_{[q_1q_2q_3q_4]}(z,\bar z)=(-1)^{\sum_{i=1}^4 j_i}g_{[-q_1,-q_2,-q_3,-q_4]}(\bar z,z).
	\ee
	This identity also holds for individual conformal blocks.
\end{itemize}

\subsection{Conformal block for identity exchange}

As a simple illustration of some of the above definitions, let us compute the expression for the conformal block for 
identity exchange in a general 3d correlator.

The identity block appears in correlation functions of the form
\be
	\<\cO_1\cO_1\cO_4\cO_4\>,
\ee
and its contribution is simply equal to
\be
	\<\cO_1(x_1,s_1)\cO_1(x_2,s_2)\cO_4(x_3,s_3)\cO_4(x_4,s_4)\>\ni \<\cO_1(x_1,s_1)\cO_1(x_2,s_2)\>\<\cO_4(x_3,s_3)\cO_4(x_4,s_4)\>.
\ee
To compute the decomposition of this block, we need to evaluate the above block in the configuration~\eqref{eq:4ptCF} using~\eqref{eq:twoptnormalization}. We find
\be
	&\<\cO_1(x_1,s_1)\cO_1(x_2,s_2)\>\<\cO_4(x_3,s_3)\cO_4(x_4,s_4)\>\nn\\
	&=c_{\cO_1}c_{\cO_4}\frac{i^{2j_1}(s_1^\a\g^\mu_{\a\b}s_2^\b x_{12,\mu})^{2j_1}}{x_{12}^{2\De_1+2j_1}}\frac{i^{2j_4}(s_3^\a\g^\mu_{\a\b}s_4^\b x_{34,\mu})^{2j_4}}{x_{34}^{2\De_4+2j_4}}.
\ee
We have
\be
	\g^{\mu,\a\b}x_{12,\mu}&=
	\begin{pmatrix}
		z & 0 \\ 0 & -\bar z
	\end{pmatrix}^{\a\b},\nn\\
	\g^{\mu,\a\b}x_{43,\mu}&=
	\begin{pmatrix}
		L-1 & 0 \\ 0 & 1-L
	\end{pmatrix}^{\a\b},
\ee
which implies
\be
	&\<\cO_1(x_1,s_1)\cO_1(x_2,s_2)\>\<\cO_4(x_3,s_3)\cO_4(x_4,s_4)\>\nn\\
	&=c_{\cO_1}c_{\cO_4}
	\frac{i^{2j_1}((s_1)_1(s_2)_1z-(s_1)_2(s_2)_2\bar z)^{2j_1}}{(z\bar z)^{\De_1+j_1}}
	{i^{2j_4}((s_3)_1(s_4)_1-(s_3)_2(s_4)_2)^{2j_4}}\nn\\
	&=\sum_{q_1,q_4}c_{\cO_1}c_{\cO_4}i^{2q_1+2q_4}\binom{2j_1}{j_1+q_1}\binom{2j_4}{j_4+q_4}
	(s_1)_1^{j_1+q_1}(s_2)_1^{j_1+q_1}
	(s_1)_2^{j_1-q_1}(s_2)_2^{j_1-q_1}\nn\\
	&\hspace{2cm}
	\x(s_3)_1^{j_4+q_4}(s_4)_1^{j_4+q_4}
	(s_3)_2^{j_4-q_4}(s_4)_2^{j_4-q_4}\x z^{-\De_1+q_1}\bar z^{-\De_1-q_1}.
\ee
We see that for the identity block, the only non-zero components are those with $q_1=q_2$ and $q_3=q_4$. These functions are
\be
	g_{[q_1q_2q_3q_4]}(z,\bar z)=c_{\cO_1}c_{\cO_4}i^{2q_1+2q_4}\binom{2j_1}{j_1+q_1}\binom{2j_4}{j_4+q_4}z^{-\De_1+q_1}\bar z^{-\De_1-q_1}.
\ee

Note that the identity block is defined without a reference to the three-point bases, and in fact, the above expressions do not need to be multiplied by OPE coefficients: they directly give the contribution of the identity operator to the four-point function.

\section{Conformal block generator~\texttt{blocks\_3d}}
\label{sec:blocks3d}

\subsection{The algorithm}
\label{sec:algo}

Approximations to conformal blocks are computed in \texttt{blocks\_3d} using residue recursion relations~\cite{Kos:2013tga,Kos:2014bka,Penedones:2015aga}, and specifically the general form of the 3-dimensional residue recursion relations derived in~\cite{Erramilli:2019njx}. In this section we briefly review these recursion relations and how they are used in~\texttt{blocks\_3d}.

The statement of residue recursion relations in $d=3$ is as follows. The conformal blocks $g^{ab}_{\De,j,I}(z,\bar z)$ are meromorphic functions of $\De\in \C$ with simple poles and known residues.\footnote{The validity of this statement is dependent on the conventions for tensor structures. It is valid for the choices discussed in this paper.} For each $j$ there is an infinite set of poles $\De_{j,i}$, labeled by an index $i$ in some index set $i\in \cP_j$. The set of these poles is independent of $a,b,I$, with the exception that some residues may vanish for special values of theses indices. The residues at these poles take the form
\be\label{eq:residuerecrel}
	g^{ab}_{\De,j,I}(z,\bar z)\sim \frac{1}{\De-\De_{j,i}}(\cL_{j,i})^a_{a'}(\cR_{j,i})^b_{b'}g^{a'b'}_{\De'_{j,i},j'_{j,i},I}(z,\bar z),
\ee
where summation over repeated indices is understood.
Here, the matrices $\cL_{j,i}$ and $\cR_{j,i}$ are $I$-independent, and in general depend on $\De_{12}$ and $\De_{43}$. The quantum numbers $\De'_{j,i},j'_{j,i}$ appearing in the right-hand side have known expressions in terms of $j$ and $i$. Importantly, we have
\be
\De'_{j,i}=\De_{j,i}+n_{j,i},
\ee
where $n_{j,i}$ are \textit{positive} integers. The quantities $\De_{j,i},n_{j,i},j'_{j,i}$ were computed in~\cite{Penedones:2015aga}. Some examples of the matrices $\cL_{j,i}$, $\cR_{j,i}$ were computed in~\cite{Kos:2013tga,Kos:2014bka,Iliesiu:2015akf,Penedones:2015aga,Dymarsky:2017xzb,Reehorst:2019pzi}, and the general closed-form expressions were derived for them in~\cite{Erramilli:2019njx}.

To make use of~\eqref{eq:residuerecrel}, the conformal blocks are separated into two factors,
\be
	g^{ab}_{\De,j,I}(z,\bar z) = (a_0 r)^\De h^{ab}_{\De,j,I}(z,\bar z),
\ee
where 
\be\label{eq:rdef}
	r=\sqrt{\r\bar\r},\quad\r = \frac{z}{(1+\sqrt{1-z})^2},\quad\bar\r = \frac{\bar z}{(1+\sqrt{1-\bar z})^2},
\ee
and $a_0$ and $b_j$ are the constants appearing in the normalization of the two-point function~\ref{eq:2ptnorm}.
The functions $h^{ab}_{\De,j,I}(z,\bar z)$ defined in this way are useful because they are holomorphic at $\De=\oo$,
\be\label{eq:hinfinity}
	h^{ab}_{\De,j,I}(z,\bar z)=h^{ab}_{\oo,j,I}(z,\bar z)+O(\tfrac{1}{\De}).
\ee
The functions $h^{ab}_{\oo,j,I}(z,\bar z)$ can be determined by solving the Casimir differential equation to leading order in $\De$, and have been computed in a general closed form in~\cite{Erramilli:2019njx}, based on the results of~\cite{Kravchuk:2017dzd}.

The relation~\eqref{eq:residuerecrel} implies the following expression for the residues of $h^{ab}_{\De,j,I}(z,\bar z)$,
\be
	h^{ab}_{\De,j,I}(z,\bar z)\sim \frac{(a_0r)^{n_{j,i}}}{\De-\De_{j,i}}(\cL_{j,i})^a_{a'}(\cR_{j,i})^b_{b'}h^{a'b'}_{\De'_{j,i},j'_{j,i},I}(z,\bar z).
\ee
Combining this with~\eqref{eq:hinfinity}, and with the knowledge that $h^{ab}_{\De,j,I}(z,\bar z)$ is meromorphic in $\De$, we find the residue recursion relation
\be\label{eq:finalresiduerec}
	h^{ab}_{\De,j,I}(z,\bar z)=h^{ab}_{\oo,j,I}(z,\bar z)+\sum_{i\in \cP_j} \frac{(a_0r)^{n_{j,i}}}{\De-\De_{j,i}}(\cL_{j,i})^a_{a'}(\cR_{j,i})^b_{b'}h^{a'b'}_{\De'_{j,i},j'_{j,i},I}(z,\bar z).
\ee

For the purposes of the numerical conformal bootstrap we need to compute an approximation to $h^{ab}_{\De,j,I}(z,\bar z)$ near $z=\bar z=\half$. It is well-known that $h^{ab}_{\De,j,I}(z,\bar z)$ has an expansion in positive integer powers of $r$ which converges quickly near this point~\cite{Hogervorst:2013sma}. The residue recursion relation~\eqref{eq:finalresiduerec} provides an efficient way of computing this power series. To see this, consider the $r^0$ term in this expansion. Since all $n_{j,i}>0$, we can completely neglect the sum over $i$. Since $h^{ab}_{\oo,j,I}(z,\bar z)$ is a known function, we can easily extract the coefficient of the $r^0$ term from its explicit expression. 

To compute the $r^1$ term, we cannot neglect the sum over $i$ anymore, but due to $n_{j,i}>0$ we only need to know the $r^0$ term of the $h^{a'b'}_{\De'_{j,i},j'_{j,i},I}(z,\bar z)$ that appear in the sum, which we have already computed. Repeating in this manner we can generate the $r$-series expansion for $h^{ab}_{\De,j,I}(z,\bar z)$ to very high orders. (In realistic applications the order can often go to $r^{80}$ or higher.)

The recursion relation~\eqref{eq:finalresiduerec} also gives a nice representation of the $\De$-dependence of $h^{ab}_{\De,j,I}(z,\bar z)$. In particular, once the $r$-series of $h^{ab}_{\De,j,I}(z,\bar z)$ has been computed to the order $r^N$, we can drop the terms with $n_{j,i}>N$ in the right-hand side of~\eqref{eq:finalresiduerec} and substitute the derivatives at $z=\bar z=\half$ of the computed series for the remaining terms. In this way, we obtain an approximation of the form
\be
\ptl_{z}^m\ptl_{\bar z}^n h^{ab}_{\De,j,I}(z,\bar z)\Big\vert_{z=\bar z=\half}\approx D_{0,j,I}^{a,b;m,n}+\sum_{i\in \cP_j, n_{j,i}\leq N} \frac{D_{i,j,I}^{a,b;m,n}}{\De-\De_{j,i}},
\ee
where $D_{i,j,I}^{a,b;m,n}\in \R$ are numbers. This can be rewritten as
\be
\ptl_{z}^m\ptl_{\bar z}^n h^{ab}_{\De,j,I}(z,\bar z)\Big\vert_{z=\bar z=\half}\approx \frac{\tl P^{a,b;m,n}_{j,I}(\De)}{\prod_{i\in \cP_j, n_{j,i}\leq N}(\De-\De_{j,i})},
\ee
where $P^{a,b;m,n}_{j,I}(\De)$ is a polynomial in $\De$ of degree equal to the number of $i\in \cP_j$ with $n_{j,i}\leq N$. For the conformal block itself we then get
\be\label{eq:polerepresentation}
\ptl_{z}^m\ptl_{\bar z}^n g^{ab}_{\De,j,I}(z,\bar z)\Big\vert_{z=\bar z=\half}\approx \frac{ b_j^{-1} (a_0r)^\De P^{a,b;m,n}_{j,I}(\De)}{\prod_{i\in \cP_j, n_{j,i}\leq N}(\De-\De_{j,i})}
\ee
for some new polynomials $P^{a,b;m,n}_{j,I}(\De)$ of degree $\mathrm{deg}\,P^{a,b;m,n}_{j,I}(\De)=\mathrm{deg}\,\tl P^{a,b;m,n}_{j,I}(\De)+m+n$. For future convenience, we also factored out the explicit dependence on $b_j$. The output of~\texttt{blocks\_3d} is essentially the polynomials $P^{a,b;m,n}_{j,I}(\De)$, with some tweaks and optimizations described below. The precise form of the output is specified in section~\ref{sec:output}.

In the rest of this section we briefly describe some more technical points about the algorithm used in~\texttt{blocks\_3d}.

\subsubsection{Block structure of residue matrices}

When using the recursion relation~\eqref{eq:finalresiduerec}, we need to multiply the known part of the power series by matrices $\cL_{j,i}$ and $\cR_{j,i}$. These matrices have size $N_3\times N_3$, where $N_3$ is the number of three-point tensor structures for three-point functions $\<\cO_1\cO_2\cO\>$ (for $\cL$) or $\<\cO_4\cO_3\cO\>$ (for $\cR$), and $\cO$ has spin $j$. In more complicated blocks $N_3$ can be relatively large. For example, for four-point stress-tensor blocks we have $N_3=25$,\footnote{This is the number of three-point tensor-structures for $\<\cO_1\cO_2\cO\>$, where $\cO_1$ and $\cO_2$ are distinct, non-conserved spin-2 operators.} which is to be compared with scalar blocks where $N_3=1$. Taking into account that the algorithmic complexity of matrix multiplication is $O(N_3^3)$, we find that just this multiplication step is $10^4$ times slower than in the case of scalar blocks.

It is, therefore, desirable to reduce $N_3$ as much as possible. In the case of conformal blocks for $\<TTTT\>$, we know that in reality, the number of three-point structures $\<TT\cO\>$ is at most $2$~\cite{Dymarsky:2017yzx}. The $N_3=25$ above comes from ignoring the permutation symmetry, space parity, and conservation properties of $\<TT\cO\>$. It thus seems to be a good idea to specialize the recursion relation~\eqref{eq:finalresiduerec} to structures which satisfy these properties. However, we do not do this in~\texttt{blocks\_3d} for the following technical reasons. 

First of all, these properties are very problem-specific and would significantly complicate the input information required by~\texttt{blocks\_3d}. Furthermore, even if we wanted to implement conservation constraints, we would have to use three-point structures that solve these constraints. Such structures depend polynomially on $\De$ and the corresponding function $h_{\De,j,I}^{ab}(z,\bar z)$ is not holomorphic at $\De=\oo$. Instead, it has a high-degree pole there (for $\<TTTT\>$ it would grow as $\De^{12}$).\footnote{It is possible to divide the three-point tensor structures by polynomials in $\De$ to cancel this pole. However, this would introduce extra poles for $\De\in \C,$, and we would need to somehow compute the residues at these poles.} This means that we would need to determine more functions at $\De=\oo$ than just $h_{\oo,j,I}^{ab}(z,\bar z)$, and general expressions for these functions are not currently available. 

Instead of relying on permutation and conservation properties of three-point structures, we use the fact observed in~\cite{Erramilli:2019njx} that the matrices $\cL_{j,i}$ and $\cR_{j,i}$ are block-diagonal in $j_{12}$ and $j_{43}$. That is, working in the $\SO(3)$ basis defined in section~\ref{sec:3ptbasis}, we have
\be
	(\cL_{j,i})^{(j_{12},j_{120})}_{(j_{12}',j'_{120})}\propto \de^{j_{12}}_{j_{12}'},\quad
	(\cR_{j,i})^{(j_{43},j_{430})}_{(j_{43}',j'_{430})}\propto \de^{j_{43}}_{j_{43}'}.
\ee
We additionally take into account the fact that these either preserve or flip (depending on $i$) the space parity of the structure. These observations allow us to split the three-point tensor structures into groups distinguished by the value of $j_{12}$ (or $j_{43}$) and space parity. In the example of $\<TTTT\>$ blocks, we split the $N_3=25$ generic structures into groups the largest of which contains only $5$ structures, which gives a significant improvement in performance over the direct application of~\eqref{eq:residuerecrel}.

Since the recursion relations for different values of $j_{12}$, $j_{43}$ and $I$ can be used completely independently, a single run of \texttt{blocks\_3d} only computes the conformal blocks
\be
g_{\De,j,I}^{(j_{12},j_{120}),(j_{43},j_{430})}(z,\bar z),\nn
\ee
with the values of $j_{12}$, $j_{43}$ and $I$ provided by the user. This enables easy parallelization of conformal block computations.

\subsubsection{Pole-shifting}
\label{sec:poleshifting}

One drawback of using the representation~\eqref{eq:polerepresentation} for the derivatives of the conformal blocks is that for large $r$-series order $N$ the polynomials $P^{a,b;m,n}(\De)$ become of high degree. These polynomials are typically used as an input to the semidefinite solver \texttt{SDPB}~\cite{Simmons-Duffin:2015qma,Landry:2019qug}, which performs slower with higher-degree polynomials. Unfortunately, taking $N$ to be relatively large is often necessary in order to get a reliable approximation for the high-order cross-ratio derivatives of the conformal blocks. 

In order to address this problem, we use the ``pole-shifting'' method first described in~\cite{Kos:2013tga} and also implemented in \texttt{scalar\_blocks}. This method introduces a new truncation parameter $\kappa$ and looks for approximations of the form
\be\label{eq:poleshiftingapprox}
\frac{P^{a,b;m,n}_{j,I}(\De)}{\prod_{i\in \cP_j, n_{j,i}\leq N}(\De-\De_{j,i})}
\approx
\frac{P'^{a,b;m,n}_{j,I}(\De)}{\prod_{i\in \cP_j, n_{j,i}\leq \kappa}(\De-\De_{j,i})},
\ee
where $P'^{a,b;m,n}_{j,I}(\De)$ are new, lower-degree polynomials. The polynomials $P'^{a,b;m,n}_{j,I}(\De)$ are chosen to ensure that the difference between the two sides of~\eqref{eq:poleshiftingapprox} is at most $O(\De^{-\lceil M/2\rceil-1})$ near $\De=\oo$ and $O((\De-\De_0(j))^{\lfloor M/2\rfloor})$ near $\De=\De_0(j)$, where $\De_0(j)$ is the unitarity bound for the given value of $j$, and $M$ is the number of poles appearing on the right-hand side.

In practice we use large $N$ and moderate $\kappa$, making sure that increasing $\kappa$ does not affect the output of the semidefinite solver.

\subsubsection{Optimizing for spinning four-point structures}
\label{sec:sbdiffs}

In~\texttt{scalar\_blocks}, the residue recursion relations are used to compute the derivatives of the blocks along $z=\bar z$ diagonal, and then the Casimir equation is used to compute the off-diagonal derivatives~\cite{ElShowk:2012ht}. This strategy is useful because the computation of off-diagonal derivatives using the Casimir equation is much more efficient than running the recursion relation multiple times.

In \texttt{blocks\_3d} this approach is not used, and the residue recursion relation is run multiple times to compute the $r$-series for all off-diagonal derivatives. This is because, in the spinning case, there are typically several four-point tensor structures $I$, and the Casimir recursion relation mixes them all together. In cases such as $\<TTTT\>$ there are hundreds of four-point tensor structures $I$. Using the Casimir recursion relation would require running the recursion step for all of them, while in practice, only the blocks for a few values of $I$~\cite{Dymarsky:2013wla,Dymarsky:2017xzb,Dymarsky:2017yzx} are needed. Computing the off-diagonal derivatives directly from the recursion relation allows us to run the code only for these few values of $I$.

\subsection{Coordinates for cross-ratios}
\label{sec:coordinates}

There are several choices of coordinates in cross-ratio space available in \texttt{blocks\_3d}: 
\begin{itemize}
	\item $z,\bar z$ coordinates are defined by the conformal frame~\eqref{eq:4ptCF} and are related to the standard $u,v$ coordinates by
	\be
		u=\frac{x_{12}^2x_{34}^2}{x_{13}^2x_{24}^2}=z\bar z,\quad 	u=\frac{x_{14}^2x_{23}^2}{x_{13}^2x_{24}^2}=(1-z)(1-\bar z).
	\ee
	The crossing-symmetric point is $z=\bar z=\half$ and crossing acts by $z\to 1-z, \bar z\to 1-\bar z$.
	\item $x,t$ coordinates are defined through $z,\bar z$ via
	\be
		x=\frac{z+\bar z-1}{2}, \quad t=\p{\frac{z-\bar z}{2}}^2.
	\ee
	The crossing-symmetric point is $x=t=0$ and crossing acts by $x\to -x$. These coordinates are useful because they make manifest the symmetry of various functions with respect to $z\leftrightarrow \bar z$.
	\item $y,\bar y$ coordinates~\cite{Mazac:2016qev} uniformize the cut $z,\bar z$-plane
	\be
		z=\frac{(1+y)^2}{2(1+y^2)},\quad \bar z=\frac{(1+\bar y)^2}{2(1+\bar y^2)}.
	\ee
	The crossing-symmetric point is $y=\bar y=0$ and crossing acts by $y\to -y,\bar y\to -\bar y$. The conformal block expansion is convergent when $|y|,|\bar y|<1$ and it is expected that the components of the extremal functional converge to finite values in these coordinates~\cite{Mazac:2016qev}.
	\item $w,s$ coordinates are the analogs of $x,t$ for $y,\bar y$,
	\be
		w=\frac{y+\bar y}{2}, \quad s=\p{\frac{y-\bar y}{2}}^2.
	\ee
	The crossing-symmetric point is $w=s=0$ and crossing acts by $w\to -w$.
\end{itemize}
It is furthermore possible to compute only the ``radial derivatives'', i.e.\ the $\ptl_x^n\ptl_t^0$ or $\ptl_w^n\ptl_s^0$ derivatives of the conformal blocks. This option is useful, for example, in cases when there are one-dimensional degrees of freedom in the system of crossing equations~\cite{Dymarsky:2017yzx,Dymarsky:2017xzb}.

\subsection{The output}
\label{sec:output}

In this section we give the precise definition of the quantities output by \texttt{blocks\_3d}. The code computes conformal blocks as defined in section~\ref{sec:defn}, in the $\SO(3)$ basis of three-point structures as defined in section~\ref{sec:3ptbasis}, and in the $q$-basis of four-point structures as defined in section~\ref{sec:4ptstructs}. The two-point functions of the exchanged operators are normalized as described in section~\ref{sec:defn}.

Let $g^{(j_{12},j_{120}),(j_{43},j_{430})}_{\De,j,[q_1q_2q_3q_4]}(z,\bar z)$ denote such conformal blocks. These blocks do not have a definite symmetry under $z\leftrightarrow \bar z$. Therefore, we define
\be
g^{(j_{12},j_{120}),(j_{43},j_{430})}_{\De,j,[q_1q_2q_3q_4],\pm}(z,\bar z)=\frac{1}{2}\p{g^{(j_{12},j_{120}),(j_{43},j_{430})}_{\De,j,[q_1q_2q_3q_4]}(z,\bar z)\pm (-1)^{\sum_{i=1}^{4} j_i}g^{(j_{12},j_{120}),(j_{43},j_{430})}_{\De,j,[-q_1,-q_2,-q_3,-q_4]}(z,\bar z)}.
\ee
The functions $g^{(j_{12},j_{120}),(j_{43},j_{430})}_{\De,j,[q_1q_2q_3q_4],\pm}(z,\bar z)$ are even under $z\leftrightarrow \bar z$ for $(+)$ sign and odd for $(-)$ sign. 

Given a user-specified choice of the coordinates $(c_1,c_2)\in\{(z,\bar z),(x,t),(y,\bar y),(w,s)\}$, we approximate the derivatives 
\be
	\ptl_{c_1}^m\ptl_{c_2}^n \left(p_\pm(c_1,c_2) g^{(j_{12},j_{120}),(j_{43},j_{430})}_{\De,j,[q_1q_2q_3q_4],\pm}(z,\bar z)\right)\Big\vert_{\text{crossing-symmetric point}},
\ee
as discussed in section~\ref{sec:algo}, in the form
\be\label{eq:derivapprox}
\approx b_j^{-1}(a_0 r_0)^\De\frac{P'^{(j_{12},j_{120}),(j_{43},j_{430});m,n}_{j,[q_1q_2q_3q_4],\pm}(\De)}{\prod_{i\in \cP_j, n_{j,i}\leq \kappa}(\De-\De_{j,i})},
\ee
where the factor $p_{\pm}(c_1,c_2)$ is introduced to ensure that we take derivatives of a smooth function. Here we choose all $p_{\pm}(c_1,c_2) = 1$ except $p_{-}(x, t) = \frac{2}{z-\bar z}$ and $p_{-}(w,s) = \frac{2}{y-\bar y}$. In addition, $r_0=3-2\sqrt 2\approx 0.1716$ is the value of $r$ at the crossing-symmetric point and $\kappa$ is the user-selected truncation order for pole-shifting (section~\ref{sec:poleshifting}). These approximations are expressed in terms of $\bx=\De-\De_0(j)$, where $\De_0(j)$ is the unitarity bound. The poles $\bx_{j,i}=\De_{j,i}-\De_0(j)$ appearing in the above approximation as well as the polynomials $P'$ (expressed in terms of $\bx$) are output. 

Given the user-specified derivative order $\L$, the values of $m,n$ which are output are determined by
\be
	m,n\geq 0, \quad m\mu(c_1)+n\mu(c_2)\leq \L_{\pm},
\ee
where the weights $\mu(c_i)$ are given by $\mu(t)=\mu(s)=2$ and for other coordinates are equal to $1$. The quantity $\L_\pm$ is defined as $\L_+=\L$ and $\L_-=\L-1$. For the coordinate choices $z,\bar z$ or $y,\bar y$ only half the derivatives are output due to the symmetry property under $z\leftrightarrow \bar z$.

The values of $j_{12},j_{43},q_i,\pm$ are fixed in a given run of \texttt{blocks\_3d}. The values of $j_{120},j_{430}$ are chosen to be compatible with the parity of the structures. For example, if the four-point tensor structure is parity-even, only the combinations of $j_{120}$ and $j_{430}$ which correspond to two parity-even or two parity-odd three-point structures are output, since the blocks vanish for other combinations.

If the exchanged operator is fermionic, i.e., $j$ is a half-integer, the polynomials $P'$ are pure imaginary. In this case, their imaginary part is output. If $j$ is an integer, the polynomials are real and are output directly.

\subsection{Implementation details}
\label{sec:implementation}

\texttt{blocks\_3d} is implemented in C++14 and uses the GMP
\cite{gmp}, Boost \cite{boost}, FMT \cite{fmt} and Eigen
\cite{eigenweb} libraries.  We repeatedly profiled the execution to
detect what parts were taking a long time and aggressively optimized
those parts.  In a small number of cases, we had to rewrite code in an
ugly fashion to reduce temporaries and memory pressure.

However, the most significant improvements came from using multiple,
thread-local caches to speed up computations.  For example,
Clebsch-Gordan coefficients must be computed many times with identical
inputs.  These caches reduce execution time by more than an order of
magnitude, but, unfortunately, they significantly increase memory use.

We also parallelized \texttt{blocks\_3d} by splitting the computation
across a user-specified number of threads.  The work proceeds in two
stages:

\begin{enumerate}
\item Compute the derivatives (\ref{eq:derivapprox}) with respect to
  $\left(r, \lambda\right)$, where $r$ is defined in~\eqref{eq:rdef} and $\l=\half\log(\r/\bar\r)$.
\item Convert the derivatives from
  $\left(r, \lambda\right)$ into the output
  coordinates (section \ref{sec:coordinates}).
\end{enumerate}

Stage 1 must finish before stage 2 can start.  Each of
those stages are independently parallelized across multiple threads.

In Stage 1, for a given derivative $\partial_{\lambda}^{n}$, we use
the recursion relation~\eqref{eq:finalresiduerec} to compute the power
series in $r$, and, from it, the $r$-derivatives $\partial_{r}^{m}$ up to 
$m+n=\Lambda$.  So all of the calculations
for a given $\partial_{\lambda}^{n}$ can be computed independently.
Symmetry under $z\leftrightarrow \bar z$
mean that we only have to compute even or odd $\l$-derivatives, so there are $\Lambda/2$ different independent
computations.

We arrange these different computations in a queue, with threads
taking work from the queue when they are ready.  So very high
$\Lambda$ calculations can benefit from larger machines.

For Stage 2, it is the different values of the spin of the internal operator
(\texttt{j-internal}) that are processed with a queue.  So the degree
of parallelization is limited by the number of elements in
\texttt{j-internal}.

For the tests we have done, the limiting factor is usually
$\Lambda/2$.  The proportion of time taken by each stage varies from
30\% to 70\%, depending on the details of the problem and the
hardware.  As long as the calculation is large enough, we see very
high utilization of all cores.

This is a fairly simple way of multithreading the computation, so we
did not encounter many problems with subtle multithreading bugs.  In
addition, we ran \texttt{blocks\_3d} under the Helgrind thread error
detector \cite{Nethercote07valgrind:a,helgrind} and found no issues.
We did find multithreaded performance problems with the Boost
multiprecision library, but that has been rectified in the latest
release of Boost (1.74).

\subsection{Correctness}

We have verified the correctness of \texttt{blocks\_3d} in several ways.

We have a separate implementation in Mathematica that, while very,
very slow, allowed us to validate all of the individual components as
well as directly compare a complete calculation for smaller test cases.

We have verified that our implementation of $h_{\oo,j,I}^{ab}(z,\bar z)$ leads to the correct leading terms of the $r$-series and that the $r$-series generated by \texttt{blocks\_3d} satisfies the quadratic conformal Casimir equation~\cite{DO2} in a number of correlation functions.  Since the Casimir equation has a unique solution for a given leading term of the $r$-series expansion, this is a robust check of the code.

We have compared the output of \texttt{blocks\_3d} to that of \texttt{scalar\_blocks} in the case of
scalar blocks, and to the blocks computed in~\cite{Dymarsky:2017yzx} in the case of $\<TTTT\>$ blocks. We found
a perfect match in both cases.

Finally, we have implemented the 3d four-fermion bootstrap (as
described in section~\ref{sec:example}) and found agreement with the
previous results~\cite{Iliesiu:2015qra}.

\section{Performance}
\label{sec:performance}

In this section we present the results of some simple performance benchmarks. In section~\ref{sec:comparison-scalar-blocks} we compare the performance of~\texttt{blocks\_3d} to that of~\texttt{scalar\_blocks}~\cite{ScalarBlocks} (for the problem of computing scalar blocks). In section~\ref{sec:spinning-examples} we describe current performance numbers for various examples of spinning blocks.

Benchmarks in this section were run on the Helios cluster at the
Institute for Advanced Study, where each node has dual 14-core (28
cores total) 64-bit Intel Xeon Broadwell processors\footnote{Intel Xeon CPU E5-2680 v4 2.40GHz} and 128GB
RAM.

\subsection{Comparison to~\texttt{scalar\_blocks}}
\label{sec:comparison-scalar-blocks}

In this section we compare the performance of \texttt{blocks\_3d} with \texttt{scalar\_blocks}~\cite{ScalarBlocks} when computing scalar conformal blocks. While both programs use the same recursion relations and pole-shifting procedures, they differ in more technical aspects, such as those discussed in section~\ref{sec:sbdiffs} and how parallelization is carried out. 

\begin{table}[t]
	\begin{center}
		\begin{tabular}{l|l|c|c}
			& program & Memory (GB) & Time (hr) \\\hline
			Set 1 &\texttt{scalar\_blocks} & 0.4 & 0.005 \\
			& \texttt{blocks\_3d} & 4 & 0.014\\\hline
			Set 2 &\texttt{scalar\_blocks} & 1.7 & 0.061 \\
			& \texttt{blocks\_3d} & 18 & 0.11
		\end{tabular}
		\\\vspace{.5cm}
		\begin{tabular}{l|c|c}
			parameter & Set 1 & Set 2 \\\hline
			$\L$      & 25 & 43\\
			\texttt{j-internal} & 0-50 & 0-88 \\
			\texttt{coordinates} & \texttt{xt}& \texttt{xt}\\
			\texttt{order}     & 80 & 90\\
			\texttt{kept-pole-order} & 30 & 40\\ 
			\texttt{precision} & 655 & 1024\\
			\texttt{num-threads} & 28 & 28 
		\end{tabular}
	\end{center}
\caption{Memory usage and total runtime for two sets of parameters for \texttt{scalar\_blocks} and \texttt{blocks\_3d}, averaged over 10 runs each.}
\label{tab:sbcomparison}
\end{table}

Table~\ref{tab:sbcomparison} shows the memory usage and runtime for two sets of parameters. The parameters for Set 1 are of medium complexity, while the parameters for Set 2 are characteristic of the hardest numerical bootstrap problems analyzed in the literature~\cite{Kos:2016ysd,Chester:2019ifh}. \texttt{blocks\_3d} is slower than \texttt{scalar\_blocks} by a factor 2-3 and uses 10 times the memory. This is to be expected since \texttt{blocks\_3d} has been optimized for a more general use case.

This makes \texttt{blocks\_3d} less efficient for computing scalar blocks,
 but not impractically so.  For example, \texttt{blocks\_3d} runs will still fit comfortably on modern cluster nodes, which typically have at least 128 GB of RAM.
If a project needs to compute spinning blocks, which take much, much
longer than 3d scalar blocks, it may simplify the workflow to only use
\texttt{blocks\_3d}.

\subsection{Spinning examples}
\label{sec:spinning-examples}

When considering the performance of block-generating code, it is important to distinguish between two cases. The first case corresponds to conformal blocks appearing in correlation functions such as $\<\f\f\f\f\>,\<\f\f TT\>,\<TJTJ\>$, etc., where $\f$ is some generic scalar operator and $T,J$ are the stress-tensor and a spin-1 conserved current, respectively.\footnote{Here we write the operators in the order $\<\cO_1\cO_2\cO_3\cO_4\>$ and it is understood that the OPE is taken between $\cO_1$ and $\cO_2$ (equivalently, $\cO_3$ and $\cO_4$).} The common trait of these correlation functions is that the differences $\De_{12}=\De_1-\De_2,\De_{43}=\De_4-\De_3$ are fixed. This could be because some operators are identical and their scaling dimensions cancel in these differences, or because some scaling dimensions are protected, such as those of $T$ and $J$. Since $\De_{12}$ and $\De_{43}$ are fixed, once the set of intermediate spins, the derivative order $\L$, and approximation-quality related parameters are selected, such blocks need to be computed only once. For this reason, this case will be called ``static.''

The second case is when $\De_{12}$ and $\De_{43}$ can vary. In this situation the blocks will need to be recomputed many times in a typical bootstrap computation, which is why we will call this case ``dynamic.'' For example, in mixed-correlator bootstrap studies of the 3d Ising CFT involving external $\s,\e$-operators, the blocks for $\<\s\e\s\e\>$ and $\<\s\e\e\s\>$~\cite{Kos:2014bka} are required. Searches over the parameter space $\De_\s,\,\De_\e$ typically require on the order of $\geq 10^2$ points. More complicated setups, such as the $\mathrm{O}(2)$ model~\cite{Chester:2019ifh} which used 3 external primary operators, require more blocks per point, and the total number of scalar blocks that need to be computed in these problems can reach $10^3-10^4$.

Since we are reviewing the performance of \texttt{blocks\_3d}, we will put it in the context of the simplest setups with spinning operators that have not yet been studied with the numerical conformal bootstrap. Since quite a few single-correlator setups have already been implemented~\cite{Rattazzi:2008pe,Iliesiu:2015qra,Iliesiu:2017nrv,Dymarsky:2017xzb,Dymarsky:2017yzx}, we focus on problems which involve a pair of external primaries.\footnote{Of these, only the mixed system involving a scalar with a spin-1 conserved current has been studied in the published literature~\cite{Reehorst:2019pzi}.} We will consider systems involving correlators of $\{\f,T\}$, $\{\psi,T\}$, as examples of relatively complicated systems,\footnote{We do not consider, for example, $\{\psi,J\}$ since it has smaller blocks than $\{\psi,T\}$. For another example, we do not consider the $\{J,T\}$ system because it only has static blocks, and the worst-case static $\<TTTT\>$ block is covered in, e.g., the $\{\f,T\}$ system.} and $\{\f,\psi\}$ as an example of a relatively simple system. Here $\f$ is a generic neutral or $\Z_2$-odd scalar, and $\psi$ is a generic Majorana fermion. For simplicity, we do not consider non-trivial global symmetries: they tend not to greatly increase the number of \textit{conformal} blocks that we need to compute, and instead simply add a layer of flavor blocks. Similarly, in counting structures, we will assume that systems preserve space parity. Ignoring parity symmetry will introduce only a constant factor change in the estimates. The systems we consider, together with their correlators and numbers of four-point tensor structures are given in table~\ref{tab:systems}.

\begin{table}[t]
\begin{center}
\begin{tabular}{c|c|c}
	system & correlator & $N_4$\\
	\hline
	$\{\f,T\}$ & $\<\f\f\f\f\>$ & 1\\
	& $\<T\f\f\f\>$ & $2\x2$\\
	& $\<TT\f\f\>$ & 3\\
	& $\<T\f T \f\>$ & $2\x2$\\
	& $\<TTT\f\>$ & $2\x 4$\\
	& $\<TTTT\>$ & 5\\
	\hline
	$\{\psi,T\}$ & $\<\psi\psi\psi\psi\>$ & 1\\
	& $\<TT\psi\psi\>$ & 6\\
	& $\<T\psi T \psi\>$ & $2\x6$\\
	& $\<TTTT\>$ & 5\\
\end{tabular}
\quad
\begin{tabular}{c|c|c}
	system & correlator & $N_4$\\
	\hline
	$\{\f,\psi\}$ & $\<\f\f\f\f\>$ & 1\\
	& $\<\psi\psi\f\f\>$ & 2\\
	& $\<\f\psi\f\psi\>$ & $2\x 2$\\
	& $\<\psi\psi\psi\psi\>$ & 4\\
	\hline
\end{tabular}
\end{center}
\caption{The systems of correlators that we consider in our performance comparison, along with the number of four-point structures $N_4$ needed for each case. The notation $a\x b$ for $N_4$ means that there are $a$ different orderings of the operators (e.g.~the orderings $\<T\f T\f\>$ and $\<T\f\f T\>$), modulo $\Z_2\x\Z_2$ permutations, for the fixed OPE channel, and each ordering has $b$ four-point tensor structures. Such orderings have the same computational complexity. We are ignoring the 1- and 0-dimensional degrees of freedom in four-point structures of conserved operators~\cite{Dymarsky:2017xzb,Dymarsky:2017yzx}, since those have a much smaller computational complexity than the 2-dimensional degrees of freedom.}
\label{tab:systems}
\end{table}

To compute blocks for a given four-point function we in general need to call \texttt{blocks\_3d} several times. A separate call is required for each ordering of the operators (modulo $\Z_2\x\Z_2$ kinematic permutations which preserve the cross-ratios \cite{Kravchuk:2016qvl}), four-point tensor structure, and for every possible choice of $j_{12}$ and $j_{43}$. The latter choice is the main determining factor for the performance of \texttt{blocks\_3d} since the blocks computed in any given run are two matrices of sizes $L_1(j_{12})\x L_1(j_{43})$ and $L_2(j_{12})\x L_2(j_{43})$,\footnote{This is true for the correlators and structures we considered in our benchmarks. Depending on how space parities align, these could instead be $L_1(j_{12})\x L_2(j_{43})$ and $L_2(j_{12})\x L_1(j_{43})$.} where for integer $j$
\be
	L_1(j)\equiv j+1,\quad L_2(j)\equiv j
\ee
and for half-integer $j$
\be
	L_1(j)\equiv L_2(j)\equiv j+\half.
\ee
Theoretically, the algorithmic complexity of the recursion step depends on these sizes as
\be\label{eq:theoreticalscaling}
	\sum_{i=1,2} L_i(j_{12})L_i(j_{43})(L_i(j_{12})+L_i(j_{43})).
\ee
This scaling describes the data in table~\ref{tab:resources} reasonably well, accounting for most of the variation in the runtimes.\footnote{To be more precise, after dividing the total user time of these runs by~\eqref{eq:theoreticalscaling}, we obtain a factor of 4 difference between the highest and lowest fractions, compared to a factor of 300 difference without dividing by~\eqref{eq:theoreticalscaling}.}

To get some sense of the performance for a given correlator, we can run~\texttt{blocks\_3d} with the maximal values of $j_{12}$ and $j_{43}$ for one choice of four-point tensor structure. When using the parameters in table~\ref{tab:parameters}, the required memory resource, and runtimes for various correlation functions are shown in table~\ref{tab:resources}. 

\begin{table}[t]
\begin{center}
	\begin{tabular}{l|c}
		parameter & value \\\hline
		$\L$      & 25\\
		\texttt{j-internal} & 0-50\\
		\texttt{coordinates} & \texttt{xt}\\
		\texttt{order}     & 80\\
		\texttt{kept-pole-order} & 30\\ 
		\texttt{precision} & 655\\
		\texttt{num-threads} & 13\\
	\end{tabular}\\
\end{center}
\caption{Parameters used in our performance comparison. We use 13 threads because the parallelism is limited by $\lceil{\L/2}\rceil$ in the current implementation. For the scaling dimension-dependent parameters \texttt{delta-12},~\texttt{delta-43} and~\texttt{delta-1-plus-2} we use the values appropriate for the correlator, assigning some generic scaling dimensions to $\f$ and $\psi$.}
\label{tab:parameters}
\end{table}

\begin{table}[t]
\begin{center}
	\begin{tabular}{c|c|c|c|c|c}
		block & $j_{12}$ & $j_{43}$ &Memory (GB) & Time (hr)\\ \hline
		$\<\f\f\f\f\>$     & 0 & 0 & 4 & 0.014 \\
		$\<\f\psi\f\psi\>$ & $\half$ & $\half$ & 7  & 0.025\\
		$\<T\f\f\f\>$      & 2 & 0 & 11 & 0.045 \\
		$\<\psi\psi\psi\psi\>$ & 1 & 1 & 15 & 0.068 \\
		$\<T\f T\f\>$      & 2 & 2 & 36 &  0.20  \\
		$\<T\psi T\psi\>$  & $\frac{5}{2}$ & $\frac{5}{2}$ & 48 & 0.62  \\
		$\<TTT\f\>$        & 4 & 2 & 62 &  0.94 \\
		$\<TTTT\>$         & 4 & 4  & 106  & 6.9 
	\end{tabular}
\end{center}
\caption{Computing resources required for one call to \texttt{blocks\_3d} for each kind of block, using the maximal values of $j_{12}, j_{43}$ and for a single choice of four-point structure, given the parameters in table~\ref{tab:parameters}.}
\label{tab:resources}
\end{table}

To estimate the total time needed to compute all conformal blocks for a given correlator, we can then multiply these runtimes by the number of four-point tensor structures, as well as sum over all possible values of $j_{12}$ and $j_{43}$ assuming the scaling in \eqref{eq:theoreticalscaling}. Specifically, if $t$ is the time it takes to run~\texttt{blocks\_3d} for the maximal allowed values $j_{12,\max}, j_{43,\max}$, then we estimate the total time $t_{tot}$ required to compute conformal blocks for the given correlator as

\be
t_{tot}=t\x N_4\x \sum_{j_{12}=j_{12,\min}}^{j_{12,\max}}\sum_{j_{43}=j_{43,\min}}^{j_{43,\max}}\frac{\sum_{i=1,2} L_i(j_{12})L_i(j_{43})(L_i(j_{12})+L_i(j_{43}))}
{\sum_{i=1,2} L_i(j_{12,\max})L_i(j_{43,\max})(L_i(j_{12,\max})+L_i(j_{43,\max}))}.
\ee
where $j_{12,\min}$ and $j_{43,\min}$ equal $0$ or $\half$, and the sums proceed in integer steps.  Taking into account the number of cores reserved for the computation, we can then get the approximate estimates shown in table~\ref{tab:cpuhours} for the total CPU time required to compute the dynamic conformal blocks in the setups mentioned above (for one fixed choice of external dimensions), and the estimates in table~\ref{tab:cpuhours_static} for computing some of the static blocks in these systems. Note that the $\<TTTT\>$ correlator gives the worst-case scenario for static blocks involving scalars, fermions, and $T$.

\begin{table}[t]
	\begin{center}
		\begin{tabular}{c|c|c}
			system & correlator & CPU hours\\
			\hline
			$\{\f,\psi\}$ & $\<\f\psi\f\psi\>$ & 1.3\\
			\hline
			$\{\f,T\}$ & $\<T\f\f\f\>$ & 3.9\\
			& $\<T\f T \f\>$ & 29\\
			& $\<TTT\f\>$ & 390\\
			\hline
			$\{\psi,T\}$ &  $\<T\psi T \psi\>$ & 300 \\
		\end{tabular}
	\end{center}
	\caption{Estimates of CPU hours needed for the computation of dynamic blocks in each system of correlators (for one fixed choice of external dimensions). The notation for the correlators is the same as in table~\ref{tab:systems}.}
	\label{tab:cpuhours}
\end{table}

\begin{table}[t]
	\begin{center}
		\begin{tabular}{c|c|c}
			system & correlator & CPU hours\\
			\hline
			$\{\f,\cdots\}$ & $\<\f\f\f\f\>$ & 0.18\\
			\hline
			$\{\psi,\cdots\}$ & $\<\psi\psi\psi\psi\>$ & 6.2\\
			\hline
			$\{T,\cdots\}$ &  $\<TTTT\>$ & 2700 \\
		\end{tabular}
	\end{center}
	\caption{Estimates of CPU hours needed for the computation of select static blocks in various systems of correlators.}
	\label{tab:cpuhours_static}
\end{table}

These numbers show that it is practical to use~\texttt{blocks\_3d} for the numerical conformal bootstrap of the systems considered in this section. Specifically, the time to compute $\<TTTT\>$ block dominates the static block computation time in all setups, and we estimate it to be on the order of $2700$ CPU hours. Furthermore, assuming that the number of points in scaling dimension space for which the dynamic blocks need to be computed is on the order of $10^2-10^3$, we see that in all cases, the dynamic blocks dominate the conformal block computation time, and is estimated in total to be around $10^3-10^5$ CPU hours depending on the problem. While this is significant, this is still below the typical computational time required to run semidefinite programming for problems of this size, which can be $10^6$ CPU hours or higher.

\section{A worked example: 3d four-fermion bootstrap}

\label{sec:example}

\subsection{Physical setup}
\label{sec:physical-setup}

In this section we apply \texttt{blocks\_3d} to an example problem of the 3d four-fermion bootstrap. That is, we impose the crossing symmetry constraints on the four-point function 
\be
	\<\psi\psi\psi\psi\>
\ee
of a single Majorana fermion $\psi$ in a parity-preserving 3d CFT. The numerical bootstrap applied to this correlator was studied in great detail in~\cite{Iliesiu:2015qra}. The goal here is mostly to demonstrate how \texttt{blocks\_3d} can be applied to an interesting physical problem. To have a concrete physical goal, we will revisit some of the features of the exclusion plots of~\cite{Iliesiu:2015qra} in the context of the recently described ``fake primary" effect~\cite{Karateev:2019pvw}.

The first step is to identify the three-point structures of the operators that appear in the $\psi\x\psi$ OPE. Consider the three-point function
\be
	\<\psi\psi\cO\>,
\ee
where the operator $\cO$ has spin $j$. Assume first that $j\geq 1$. According to the discussion in section~\ref{sec:3ptbasis}, the following $q$-basis three-point tensors structures are possible for this three-point function,\footnote{In this section we use a shorthand notation where we denote the structure $\<\cO_1\cO_2\cO_3\>^{[q_1q_2q_3]}$ simply by its label $[q_1q_2q_3]$.}
\be
	[\thalf,\thalf,-1]^\pm, [\thalf,-\thalf,0]^\pm,
\ee
where we have defined
\be
	[q_1q_2q_3]^\pm\equiv [q_1q_2q_3]\pm[-q_1,-q_2,-q_3].
\ee
We need to additionally impose the requirements of permutation symmetry between the first two operators as well as the parity constraints. These structures transform under the (12) permutation (see section~\ref{sec:3ptbasis}) as
\be
	[\thalf,\thalf,-1]^\pm\to \pm (-1)^{j-1}[\thalf,\thalf,-1]^\pm,\quad 
	[\thalf,-\thalf,0]^\pm\to (-1)^{j-1}[\thalf,-\thalf,0]^\pm.
\ee
Note that since $\psi$ is a fermion we need anti-symmetric structures. We then find the following allowed structures for various types of operators that appear in $\psi\x\psi$ OPE:
\begin{itemize}
	\item Parity-even, even $j$: 
	\be
		[\thalf,-\thalf,0]^+,\, [\thalf,\thalf,-1]^+.
	\ee
	\item Parity-even odd-$j$ operators are forbidden.
	\item Parity-odd, even $j$:
	\be
	[\thalf,-\thalf,0]^-.
	\ee
	\item Parity-odd, odd $j$:
	\be
	[\thalf,\thalf,-1]^-.
	\ee
\end{itemize}
For $j=0$ the only difference is that we have to remove the second structure for the parity-even even-$j$ operators. The OPE coefficients corresponding to these structures are pure imaginary.

We now need to determine the four-point tensor structures and the corresponding crossing equations. In principle there are $2^4=16$ four-point tensor structures, corresponding, in the $q$-basis, to all possible choices of the four $q_i\in \{-\thalf,\thalf\}$. Of these structures, $8$ are parity-even. After symmetrizing under the $\Z_2\x\Z_2$ kinematic permutations (i.e., those permutations of the four operators which do not change the cross-ratios)~\cite{Kravchuk:2016qvl}, there are $5$ allowed structures, which take the form
\be
	\<++++\>&=[\thalf,\thalf,\thalf,\thalf],\label{eq:psi4pt1}\\
	\<----\>&=[-\thalf,-\thalf,-\thalf,-\thalf],\\
	\<++--\>&=[\thalf,\thalf,-\thalf,-\thalf]+\tfrac{\bar z}{z}[-\thalf,-\thalf,\thalf,\thalf],\\
	\<+-+-\>&=[\thalf,-\thalf,\thalf,-\thalf]+[-\thalf,\thalf,-\thalf,\thalf],\\
	\<-++-\>&=[-\thalf,\thalf,\thalf,-\thalf]+\tfrac{1-\bar z}{1-z}[\thalf,-\thalf,-\thalf,\thalf],\label{eq:psi4pt5}
\ee
where we use the notation $\<\cdots\>$ to denote the symmetrized structures. The four-point function can be expanded as
\be
\<\psi\psi\psi\psi\>=&\<++++\>g_{[\half,\half,\half,\half]}(z,\bar z)+
\<----\>g_{[-\half,-\half,-\half,-\half]}(z,\bar z)+\nn\\
&\<++--\>g_{[\half,\half,-\half,-\half]}(z,\bar z)+
\<+-+-\>g_{[\half,-\half,\half,-\half]}(z,\bar z)+\nn\\
&\<-++-\>g_{[-\half,\half,\half,-\half]}(z,\bar z).
\ee
The crossing equations in terms of these structures can then be written as~\cite{Kravchuk:2016qvl}
\be
g_{[\half,\half,\half,\half]}(z,\bar z)&=g_{[\half,\half,\half,\half]}(1-z,1-\bar z),\\
g_{[-\half,-\half,-\half,-\half]}(z,\bar z)&=g_{[-\half,-\half,\half,\half]}(1-z,1-\bar z),\\
g_{[\half,\half,-\half,-\half]}(z,\bar z)&=g_{[-\half,\half,\half,-\half]}(1-z,1-\bar z),\\
g_{[-\half,\half,-\half,\half]}(z,\bar z)&=g_{[-\half,\half,-\half,\half]}(1-z,1-\bar z).
\ee
We now take the derivatives of these structures near $z=\bar z=\half$ to obtain the basis of crossing equations for the numerical bootstrap.
In fact, there is a small subtlety related to the degeneration of the four-point $q$-basis on the line $z=\bar z$, and some derivatives need to be omitted. We refer the reader to appendix A of~\cite{Kravchuk:2016qvl} for a detailed discussion, where the present example is worked out.

The functions $g_{[q_1,q_2,q_2,q_4]}(z,\bar z)$ have the conformal block expansions\footnote{Note that $\l_{\psi\psi\cO}$ are pure imaginary.}
\be
	g_{[q_1,q_2,q_2,q_4]}(z,\bar z) &= \sum_{\cO} \l_{\psi\psi\cO,a} \l_{\psi\psi\cO,b}\,g_{\De_\cO,j_\cO,[q_1,q_2,q_2,q_4]}^{ab}(z,\bar z),
\ee
where the three-point indices $a,b$ label the three-point structures described above. Plugging these equations into the above crossing equations, we obtain the sum rules which can be analyzed using standard numerical bootstrap techniques.

\subsection{Translating to~\texttt{blocks\_3d} conventions}

Since each individual conformal block contributing to $\<\psi\psi\psi\psi\>$ should be decomposable into the structures~\eqref{eq:psi4pt1}-\eqref{eq:psi4pt5}, we find that, for example,
\be
g^{ab}_{\De,j,[-\thalf,-\thalf,\thalf,\thalf]}(z,\bar z)=(\bar z/z)g^{ab}_{\De,j,[\thalf,\thalf,-\thalf,-\thalf]}(z,\bar z).
\ee
This implies that
\be
g^{ab}_{\De,j,[-\thalf,-\thalf,\thalf,\thalf],+}(z,\bar z)=\frac{z+\bar z}{2z}g^{ab}_{\De,j,[-\thalf,-\thalf,\thalf,\thalf]}(z,\bar z),
\ee
so computing

\be
g^{ab}_{\De,j,[-\thalf,-\thalf,\thalf,\thalf]}(z,\bar z)
\ee
is equivalent to computing\footnote{Equivalently, we can use $g^{ab}_{\De,j,[-\thalf,-\thalf,\thalf,\thalf],-}(z,\bar z)$ instead.}
\be
g^{ab}_{\De,j,[-\thalf,-\thalf,\thalf,\thalf],+}(z,\bar z).
\ee
Therefore, in order to compute the required conformal blocks, we run \texttt{blocks\_3d} for the following choices of $q$-basis four-point structures and sign $\pm$ (see section~\ref{sec:output}),
\be
	&\{[\thalf,\thalf,\thalf,\thalf],+\},\\
	&\{[\thalf,\thalf,\thalf,\thalf],-\},\\
	&\{[\thalf,\thalf,-\thalf,-\thalf],+\},\\
	&\{[\thalf,-\thalf,\thalf,-\thalf],+\},\\
	&\{[-\thalf,\thalf,\thalf,-\thalf],+\}.
\ee

It remains to express the blocks $g^{ab}_{\De,j,[q_1q_2q_3q_4],\pm}(z,\bar z)$ with $a,b$ labeling the $q$-basis structures defined in the previous section in terms of the blocks $g^{(j_{12},j_{120}),(j_{43},j_{430})}_{\De,j,[q_1q_2q_3q_4],\pm}(z,\bar z)$ which are computed by \texttt{blocks\_3d}. For this we need to express the $q$-basis tensor structures in terms of the $\SO(3)$ basis structures. Suppose the coefficients are related by matrices $\cM^a_{j,(j_{12},j_{120})}$, then we have
\be
	&g^{ab}_{\De,j,[q_1q_2q_3q_4],\pm}(z,\bar z)\nn\\
	&=\sum_{j_{12}=0,1\atop j_{43}=0,1}\sum_{j_{120}=|j-j_{12}|}^{j+j_{12}}\sum_{j_{430}=|j-j_{43}|}^{j+j_{43}}
	\cM^a_{j,(j_{12},j_{120})}
	\cM^b_{j,(j_{43},j_{430})}
	g^{(j_{12},j_{120}),(j_{43},j_{430})}_{\De,j,[q_1q_2q_3q_4],\pm}(z,\bar z).
\ee
Note that there are 4 pairs of $j_{12},j_{43}$ entering the above sums. This means that we need to make 4 calls to \texttt{blocks\_3d} for each choice of all other parameters.

It remains to determine the matrices $\cM^a_{j,(j_{12},j_{120})}$, i.e.\ to express the $q$-basis structures in terms of the $\SO(3)$-basis structures. To keep the exposition short, we do this for a single $q$-basis structure $[\thalf,\thalf,-1]^{+}$.
We have 
\be
	[\thalf,\thalf,-1]^{+}=[\thalf,\thalf,-1]+[-\thalf,-\thalf,1].
\ee
For $[\thalf,\thalf,-1]$ we can write, interpreting it as the value of the $q$-basis structure in the configuration~\eqref{eq:qdefn}, and according to definition~\eqref{eq:so2defn}
\be
	[\thalf,\thalf,-1]=(-1)^{1-j}\binom{2j}{j-1}^{-\half}|\thalf,\thalf;\thalf,\thalf;j,-1\>.
\ee
Using~\eqref{eq:conversionSO3toq-2} we have
\be
	|\thalf,\thalf;\thalf,\thalf;j,-1\>&=\<0,1|\thalf,\thalf,\thalf;\thalf,\thalf\>|0,1;j,-1\>+\<1,1|\thalf,\thalf,\thalf;\thalf,\thalf\>|1,1;j,-1\>\nn\\
	&=|1,1;j,-1\>,
\ee
where we plugged in the values of the Clebsch-Gordan coefficients. According to~\eqref{eq:conversionSO3toq-1} we have
\begin{eqnarray}
|1,1;j,-1\> & = & \<j-1,0|1,1;j,-1\>|1,j-1\>+\<j,0|1,1;j,-1\>|1,j\>\nn\\
 &  & +\<j+1,0|1,1;j,-1\>|1,j+1\>\nn\\
 & = & \half\sqrt{\frac{j+1}{j+\half}}|1,j-1\>+\frac{1}{\sqrt{2}}|1,j\>+\half\sqrt{\frac{j}{j+\half}}|1,j+1\>,
\end{eqnarray}
and so altogether we find
\be
[\thalf,\thalf,-1]=(-1)^{1-j}\binom{2j}{j-1}^{-\half}\p{
	\half\sqrt{\frac{j+1}{j+\half}}|1,j-1\>+\frac{1}{\sqrt{2}}|1,j\>+\half\sqrt{\frac{j}{j+\half}}|1,j+1\>
}
\ee
Analogously, for $[-\thalf,-\thalf,1]$ we find
\be
[-\thalf,-\thalf,1]=(-1)^{1-j}\binom{2j}{j-1}^{-\half}\p{
-\half\sqrt{\frac{j+1}{j+\half}}|1,j-1\>+\frac{1}{\sqrt{2}}|1,j\>-\half\sqrt{\frac{j}{j+\half}}|1,j+1\>
}.
\ee
Therefore,
\be
[\thalf,\thalf,-1]^+=(-1)^{1-j}\sqrt{2}\binom{2j}{j-1}^{-\half}|1,j\>.
\ee
Recall that the left-hand side was interpreted above as the value of $q$-basis structure in configuration~\eqref{eq:qdefn}, and that $\SO(3)$ basis is defined by~\eqref{eq:SO3defn} in the same configuration. This means that we can directly read this equation as the relation between $\SO(3)$ and $q$-basis three-point tensor structures. The relations for other $q$-basis structures can be obtained analogously.

This completes the reduction of conformal blocks that are needed for our analysis to the blocks computed by \texttt{blocks\_3d}.

\subsection{Results}

\begin{figure}[t]
	\centering
	\begin{subfigure}{0.47\textwidth}
		\includegraphics[width=\textwidth]{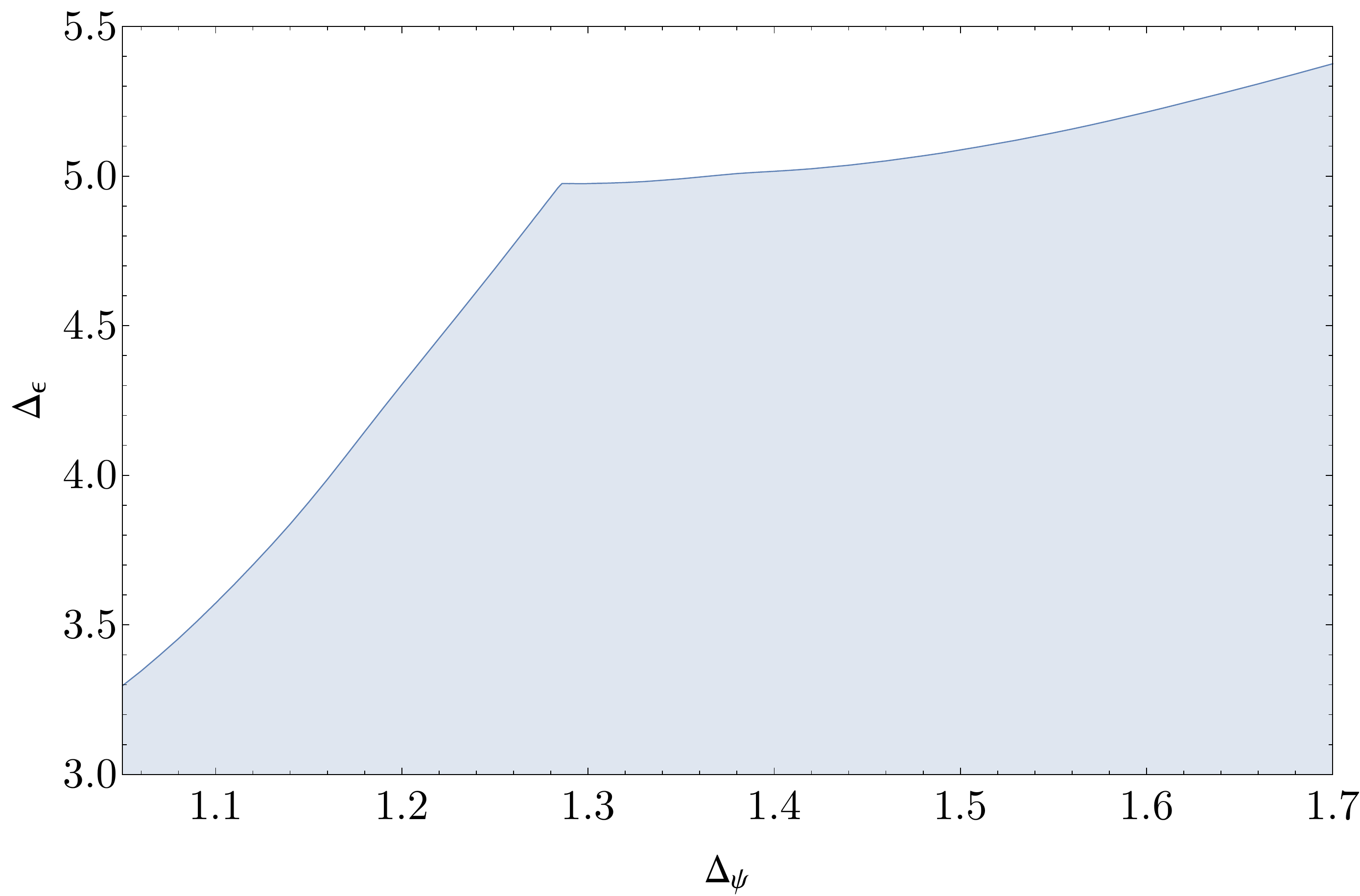}
	\end{subfigure}
	~
	\begin{subfigure}{0.47\textwidth}
		\includegraphics[width=\textwidth]{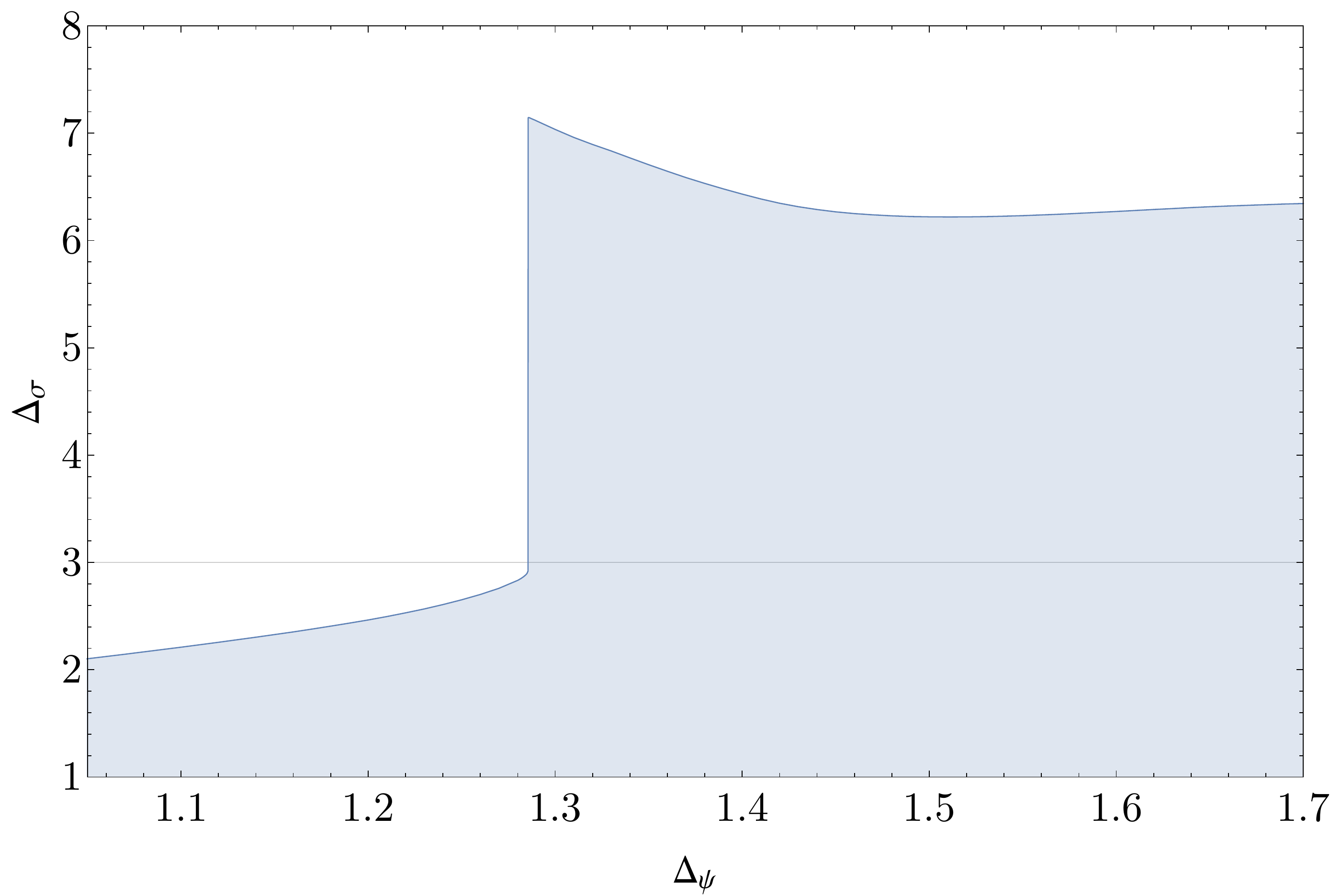}
	\end{subfigure}
	\caption{Left: the bound on the gap $\De_\e$ in the parity-even scalar sector. Right: the bound on the gap $\De_\s$ in the parity-odd scalar sector. Both plots were computed at $\L=27$.}
	\label{fig:simplegaps}
\end{figure}

We first reproduce the two simple bounds originally computed in~\cite{Iliesiu:2015qra}. These are the bounds on the gaps in scalar parity-even ($\De_\e$) and parity-odd sectors ($\De_\s$), as functions of the scaling dimension of the fermion $\De_\psi$. The results are shown in figure~\ref{fig:simplegaps}. These plots, as well as all other plots in this section, were computed at $\L=27$.\footnote{$\L$ is the upper cutoff on the total order of derivatives of the crossing equations. The other relevant numerical parameters are given in appendix~\ref{app:params}.} These results are consistent with those of~\cite{Iliesiu:2015qra} (they do not exactly coincide since we used a slightly higher $\L$) and show two prominent features. 

The bound on $\De_\e$ has a pronounced kink somewhere in the interval
\be
\De_\psi\in[1.284,1.288],
\ee
while the bound on $\De_\s$ has a sharp jump somewhere in the interval\footnote{We determined the location of the jump more precisely than that of the kink only because we study the structure of the jump in more detail below.}
\be
\De_\psi\in[1.2855250,1.2855275].
\ee
Note that these ranges are valid for the given $\L=27$, and may shift at higher derivative truncation orders. Nevertheless, these ranges clearly overlap, and it was conjectured in~\cite{Iliesiu:2015qra} that these features correspond to an actual CFT.

In~\cite{Iliesiu:2015qra}, these features, and in particular the jump in $\De_\s$, were compared to similar features in the bootstrap of the 3d Ising CFT~\cite{Kos:2014bka}. Furthermore, since the analysis of~\cite{Iliesiu:2015qra}, jumps similar to that in $\De_\s$ have been observed in the 3d fermion bootstrap with global symmetries~\cite{Iliesiu:2017nrv} and in the 4d fermion bootstrap~\cite{Karateev:2019pvw}. A common trait of all these jumps is that the jump happens when the bound approaches the number of spacetime dimensions from below. For example, in the present case, $\De_\s$ is somewhat close to 3 just to the left of the jump in figure~\ref{fig:simplegaps}.

In~\cite{Karateev:2019pvw}, the jumps in the 4d fermion bootstrap and 3d Ising bootstrap~\cite{Kos:2014bka} were shown to be due to the ``fake primary'' effect.\footnote{Importantly, however, this does not invalidate any of the 3d Ising bootstrap results~\cite{Kos:2014bka}. Moreover, there is an even sharper physical jump that remains even after the fake primary effect is removed.} We refer the reader to~\cite{Karateev:2019pvw} for a detailed general explanation of this effect. In the setup of this paper, the statement is that the exchanges of parity-odd spin-1 operators $V$ very close to the spin-1 unitarity bound $\De_V=2$ give exactly the same contribution to the four-point function $\<\psi\psi\psi\psi\>$ as exchanges of parity-odd scalars with dimension $\De_\s=3$. Thus, unless we impose a gap on $\De_V$ above $\De_V=2$, we are effectively allowing an isolated parity-odd scalar contribution at $\De_\s=3$, the ``fake primary.'' This has no effect on numerics while the gap in $\De_\s$ is below $3$, since then this isolated contribution is a part of the continuum of other allowed contributions, but it becomes important as soon as the gap in $\De_\s$ crosses $3$. In effect, in such a situation, we are bounding the gap to the second parity-odd scalar, assuming that the first parity-odd scalar is at $\De_\s=3$. This contributes to a discontinuity in the bound on $\De_\s$.

This section aims to analyze the jump in $\De_\s$ observed in figure~\ref{fig:simplegaps} in the context of the fake primary effect. While it is clear that in our setup this jump is at least partly due to the fake primary effect (because the bound on $\De_\s$ goes from below 3 to above 3), we would like to know whether there is some underlying CFT, as in the case of the 3d Ising bootstrap.

There are several indications, some seen already in the results of~\cite{Iliesiu:2015qra}, which suggest that the fake primary effect is not the primary cause of the jump. First and foremost, there is a kink observed in the bound on $\De_\e$ at the same value of $\De_\psi$. When the bound on $\De_\e$ is computed, no assumptions are made about the parity-odd scalar sector, and the fake primary is hidden in the continuum of allowed contributions. Therefore, it does not immediately affect the bound on $\De_\e$. The fact that the bound on $\De_\e$ displays a kink distinguishes the current setup from the setups in~\cite{Iliesiu:2017nrv,Karateev:2019pvw}, where the jumps seem to be only due to the fake primary effect.  Instead, it appears more similar to the situation with the 3d Ising bounds, where there is a physical theory under the jump.

\begin{figure}[t]
	\centering
	\begin{subfigure}{0.48\textwidth}
		\includegraphics[width=\textwidth]{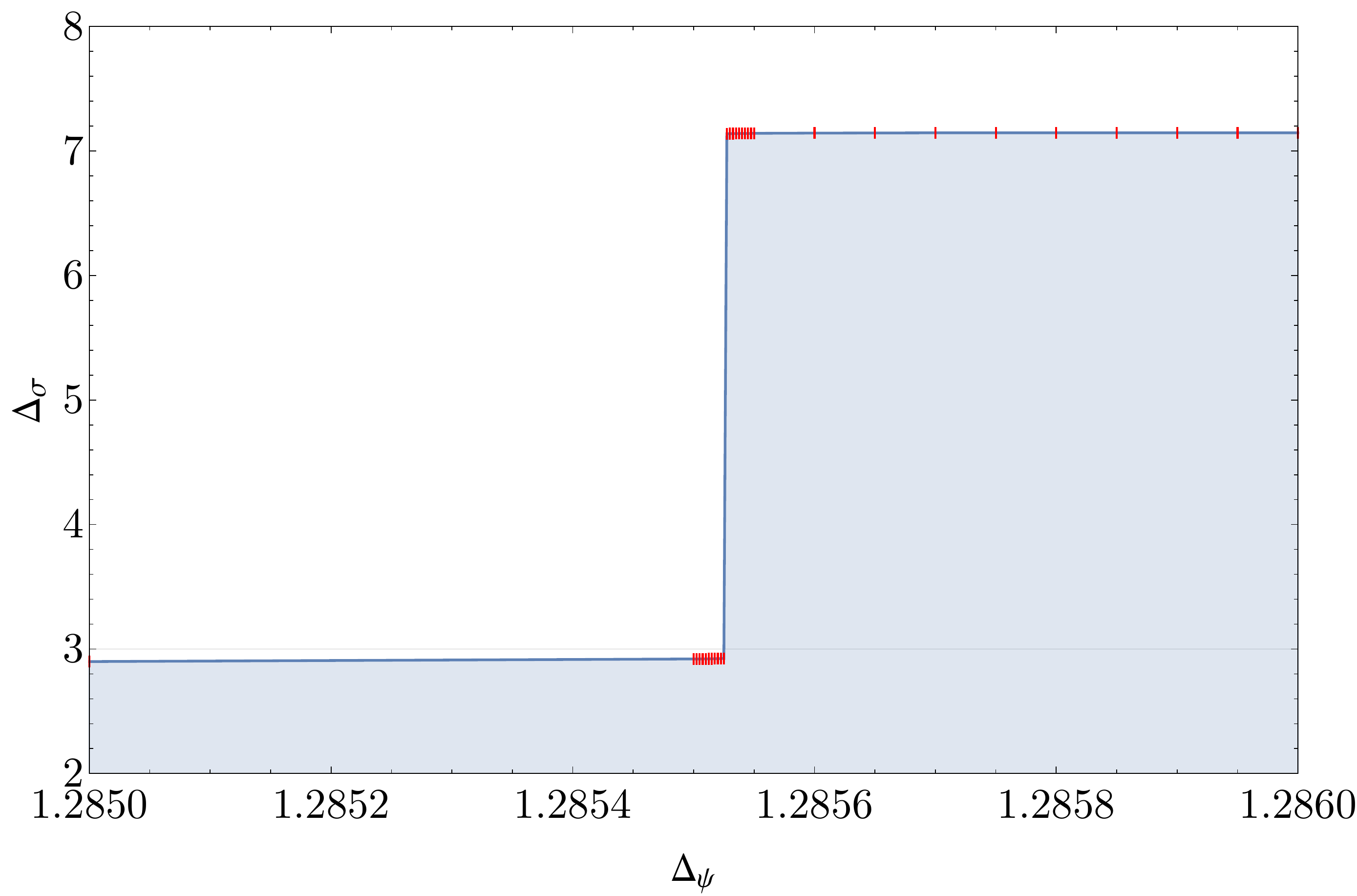}
	\end{subfigure}~
	\begin{subfigure}{0.5\textwidth}
		\includegraphics[width=\textwidth]{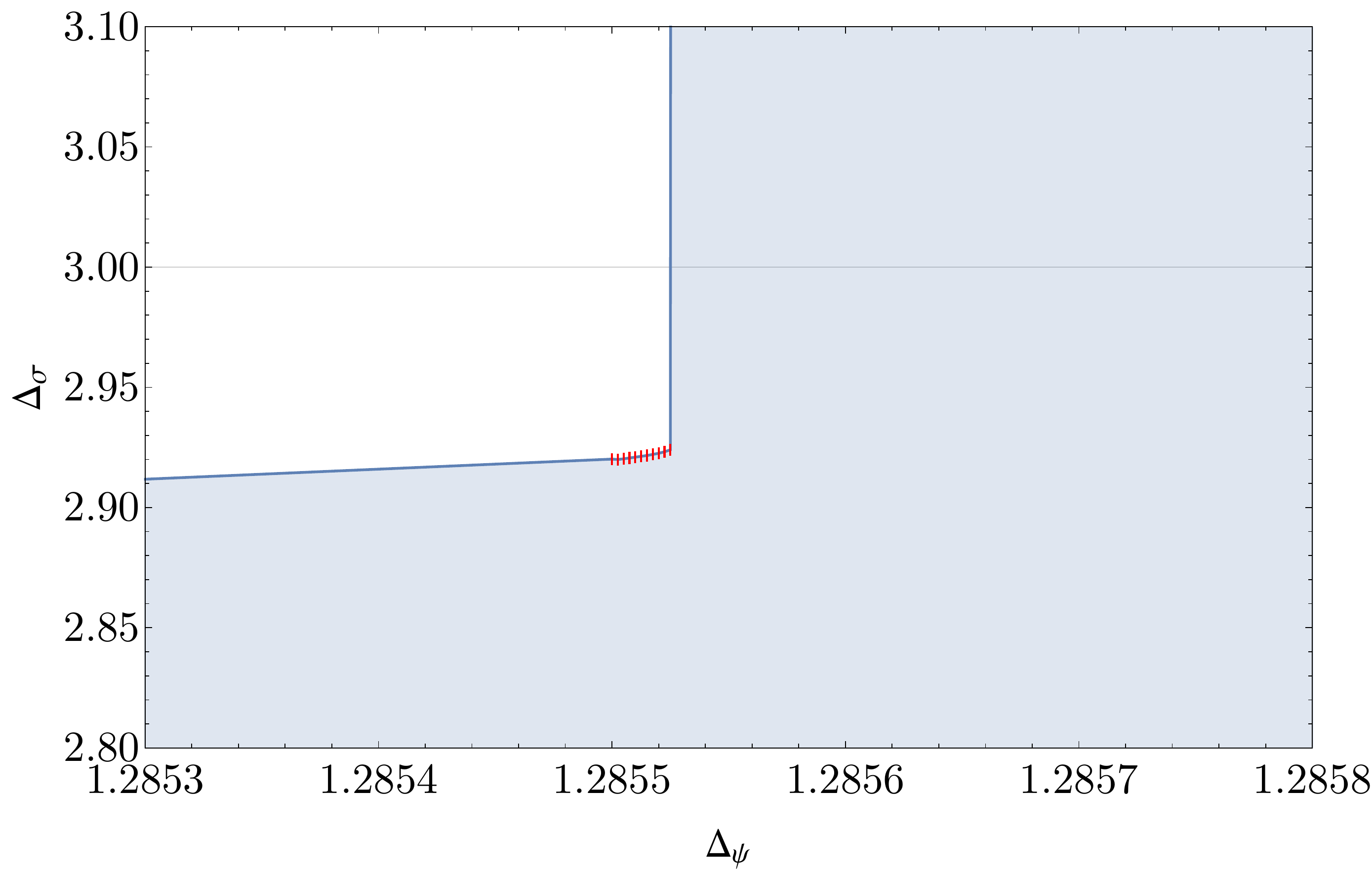}
	\end{subfigure}
	\caption{Zoom-in of the bound on $\De_\s$ near the jump at $\L=27$. The red short vertical lines show the positions of sample points in $\De_\psi$ and are not error-bars.}
	\label{fig:nearjump}
\end{figure}

Furthermore, the jump in the $\De_\s$ bound appears to start below $\De_\s=3$, which is another distinguishing feature of the jumps in 3d Ising bounds~\cite{Kos:2014bka}. To verify this, we computed the bound on $\De_\s$ over a fine grid of $\De_\psi$ values near the jump, with the results shown in figure~\ref{fig:nearjump}. These plots strongly suggest that the discontinuity starts at $\De_\s=2.924(1)$. (Again, this number is for $\L=27$.) Since we are only computing the bound at a discrete set of values $\De_\psi$, we cannot logically exclude the possibility that the true discontinuity starts at $\De_\s=3$ and is entirely due to the fake primary effect. However, in that case, there must still exist an extremely pronounced continuous feature in the plot leading up to $\De_\s=3$ just to the left of the discontinuity. This should be contrasted with the jumps observed in~\cite{Karateev:2019pvw} and~\cite{Iliesiu:2017nrv}, where the bound is perfectly smooth up to exactly the fake primary threshold, at which point it jumps.

\begin{figure}[t]
	\centering
	\begin{subfigure}{0.8\textwidth}
		\includegraphics[width=\textwidth]{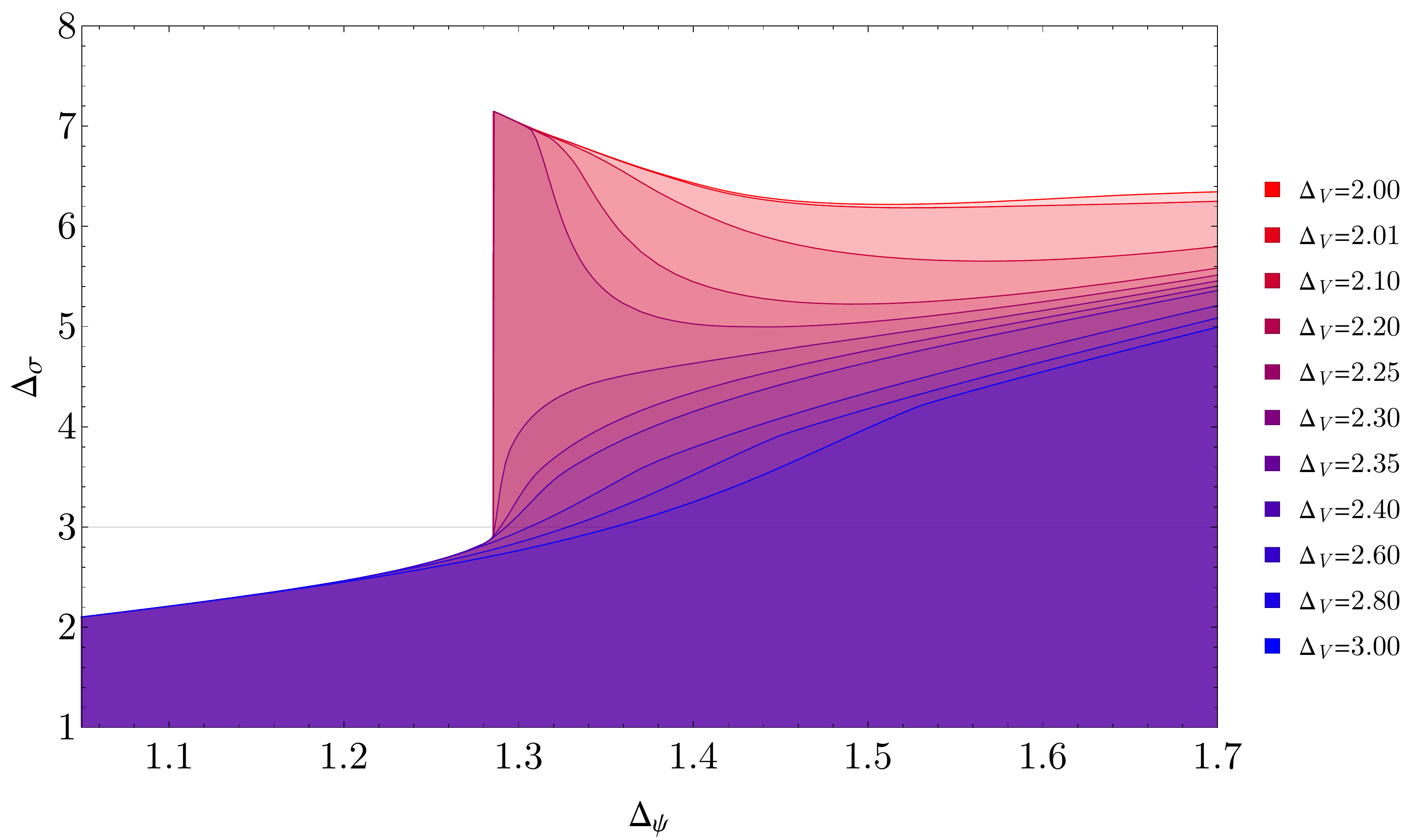}
	\end{subfigure}
	\caption{The bound on the gap $\De_\s$ in the parity-odd scalar sector at $\L=27$ for various values of the gap $\De_V\in [2,3]$. The gaps are listed on the right in the same order as the curves appear in the plot, top to bottom. The jump disappears between $\De_V=2.25$ and $\De_V=2.30$.}
	\label{fig:vectorgapsodd}
\end{figure}

The work in~\cite{Karateev:2019pvw} found that the fake primary contribution to the jump can be removed in the 3d Ising model by imposing a gap above the unitarity bound in the $\Z_2$-odd vector sector, the role of which in our setup is played by the parity-odd vectors $V$. In figure~\ref{fig:vectorgapsodd}, we show how the bound on $\De_\s$ is affected by the gaps $\De_V$ imposed on such operators. We see that the jump persists up to at least $\De_V=2.25$. The way the plot near the jump changes with $\De_V$ is somewhat different from what was observed in~\cite{Karateev:2019pvw} for the 4d fermion bootstrap, where the jumps were concluded to be likely entirely due to the fake primary effect.  But it is hard to draw sharp conclusions from this comparison.

\begin{figure}[t]
	\centering
	\begin{subfigure}{0.8\textwidth}
		\includegraphics[width=\textwidth]{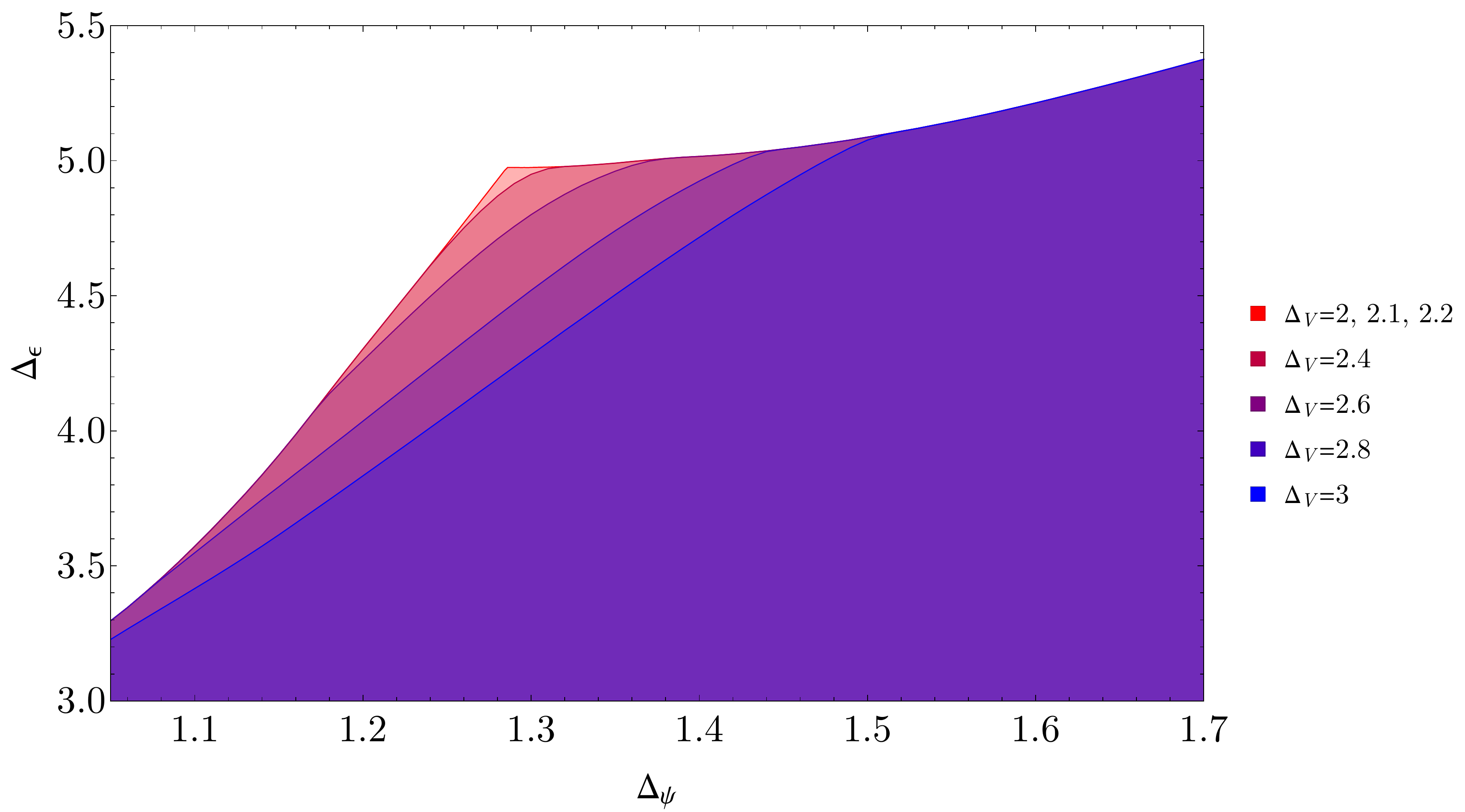}
	\end{subfigure}
	\caption{The bound on the gap $\De_\e$ in the parity-even scalar sector at $\L=27$ for various values of the gap $\De_V\in [2,3]$. The gaps are listed on the right in the same order as the curves appear in the plot, top to bottom.}
	\label{fig:vectorgapseven}
\end{figure}

It is, however, instructive to compare figure~\ref{fig:vectorgapsodd} to figure~\ref{fig:vectorgapseven}, where the bound on $\De_\e$ is plotted for various choices of $\De_V$. From figure~\ref{fig:vectorgapseven} we see that the bound is essentially independent of $\De_V$ for $\De_V\in [2.0,2.2]$, and starts to change roughly at the same time as the jump disappears. We have additionally checked that if we sit near the kink at $\{\De_{\psi}, \De_{\e}\} = \{1.286, 4.974\},$ then the maximal parity-odd spin-1 dimension is $\De_V < 2.29$ and the parity-even spin-2 gap must be smaller than $\De_T < 3.004$. It thus seems to be a consistent scenario that the jump in $\De_\s$ and the kink in $\De_\e$ are both due to a local CFT which contains a parity-odd vector operator of dimension $\De_V \approx 2.3$ as well as a stress-energy tensor with $\Delta_T = 3$.

The evidence discussed in this section appears to be inconsistent with the features in figure~\ref{fig:simplegaps} being solely explained by the fake primary effect, but is so far consistent with the existence of a local CFT with $\De_\psi\approx 1.3$, $\De_\e\approx 5$, $3\lesssim \De_\s\lesssim 7$, and a parity-odd vector operator of dimension $\De_V\approx 2.3$. It would be interesting to further explore and constrain this hypothetical CFT.

\section{Conclusions}
\label{sec:conclusions}

Introducing a general software tool for computing spinning 3d conformal blocks should mark the beginning of a new era for the numerical conformal bootstrap. In particular, \texttt{blocks\_3d} will enable the study of large systems of bootstrap equations involving external spinning operators, including fermions, global symmetry currents, and the stress tensor. In turn, this should allow for the computation of new bootstrap bounds and islands, leading to rigorous determinations of observables in physically-interesting CFTs.

An immediate future direction is to apply \texttt{blocks\_3d} to perform bootstrap computations in systems of mixed correlators containing fermions and scalars, building on the bounds obtained in~\cite{Iliesiu:2015qra, Iliesiu:2017nrv}. We expect that such a system will lead to additional constraints on the CFT data of the Gross-Neveu-Yukawa models. It may also help us explore the nature of the hypothetical ``dead-end" CFT which may underlie the kink/jump appearing in~\cite{Iliesiu:2015qra, Iliesiu:2017nrv}, in the bounds from fermion four-point functions.

It will also be interesting to perform new bootstrap computations using systems of correlators containing the stress tensor, building on the general bounds from stress-tensor four-point functions obtained in~\cite{Dymarsky:2017yzx}. In addition to allowing access to CFT observables connected to the stress tensor (e.g., three-point coefficients $\<TT\cO\>$), such systems should also help to produce a more refined map of the general space of 3d CFTs.

Another direction is to study mixed correlators containing non-Abelian currents (building on~\cite{Dymarsky:2017xzb,Reehorst:2019pzi} and the supersymmetric generalizations~\cite{Berkooz:2014yda,Li:2017ddj,Lin:2019vgi}), together with operators charged under their global symmetries. Such systems will allow for the study of whether information about current three-point coefficients can be used to help isolate 3d CFTs. They can also serve as a prototype for studying whether inputting information about 't Hooft anomalies can help isolate interesting non-supersymmetric 4d CFTs such as the conformal window of QCD. Additionally, they can be used to explore whether such correlators can be effectively used to forbid the global symmetry enhancements that affect the structure of numerous bootstrap bounds~\cite{Poland:2011ey,Nakayama:2017vdd,Li:2018lyb,Li:2020bnb,Li:2020tsm}. 

Overall, we are optimistic about the future of the numerical bootstrap. With the recent development of \texttt{SDPB 2}~\cite{Landry:2019qug}, and now the introduction of \texttt{blocks\_3d}, a plethora of new bootstrap problems involving external spinning operators should now become tractable. We expect that there is still much low-hanging fruit to be picked from these systems and that the conformal bootstrap will reveal new surprises for many years to come.

\section*{Acknowledgements}

We thank Soner Albayrak, Zhijin Li, and Emilio Trevisani for discussions. WL and DSD are supported by Simons Foundation grant 488657 (Simons Collaboration on the Nonperturbative Bootstrap). DSD is also supported by a DOE Early Career Award under grant DE-SC0019085. DP is supported by Simons Foundation grant 488651 (Simons Collaboration on the Nonperturbative Bootstrap) and DOE grants DE-SC0020318 and DE-SC0017660. LVI is supported in part by the Simons Collaboration on the Nonperturbative Bootstrap, a Simons Foundation Grant with No. 488653, and by the Simons Collaboration on Ultra-Quantum Matter, a Simons Foundation Grant with No. 651440.  PK is supported by DOE grant DE-SC0009988 and the Adler Family Fund at the Institute for Advanced Study. Computations in this work were performed on the Caltech High Performance Cluster, partially supported by a grant from the Gordon and Betty Moore Foundation, on the Yale Grace computing cluster, supported by the facilities and staff of the Yale University Faculty of Sciences High Performance Computing Center, and on the Institute for Advanced Study Helios cluster.

\appendix

\section{Code Availability}

The software \texttt{blocks\_3d} is freely available from Gitlab at

\url{https://gitlab.com/bootstrapcollaboration/blocks_3d}\\
The work presented here was computed by the latest current version,
which has the Git commit hash
\begin{lyxcode}
e37e972f5f19befa1158754ee9570c7b6a1c5913
\end{lyxcode}

\section{Details on numerics}
\label{app:params}
The computations described in section~\ref{sec:example} of this paper used the parameters given in table~\ref{tab:sdpbparams} for \texttt{SDPB}~\cite{Simmons-Duffin:2015qma,Landry:2019qug} and \texttt{blocks\_3d}.
\begin{table}[h]
	\begin{center}
		\begin{tabular}{l|c}
			parameter & value \\\hline
			$\L$ & 27\\
			spins & 0-50\\
			\texttt{kept-pole-order} & 20\\ 
			\texttt{order}     & 60\\
			\texttt{precision} & 768\\
			\texttt{dualityGapThreshold} & $10^{-30}$\\
			\texttt{primalErrorThreshold} & $10^{-200}$\\
			\texttt{dualErrorThreshold} & $10^{-200}$\\
			\texttt{findPrimalFeasible} & \texttt{false}\\
			\texttt{findDualFeasible} & \texttt{false}\\
			\texttt{detectPrimalFeasibleJump} & \texttt{true}\\
			\texttt{detectDualFeasibleJump} & \texttt{true}\\
			\texttt{initialMatrixScalePrimal} & $10^{50}$\\
			\texttt{initialMatrixScaleDual} & $10^{50}$\\
			\texttt{feasibleCenteringParameter} & $0.1$\\
			\texttt{infeasibleCenteringParameter} & $0.3$\\
			\texttt{stepLengthReduction} & $0.7$\\
			\texttt{maxComplementarity} & $10^{130}$
		\end{tabular}
	\end{center}
	\caption{Parameters used for the numerical computations in this paper.}
	\label{tab:sdpbparams}
\end{table}

\section{Conventions}
\label{app:spinorconv}

The Lorentz group in $d= 3$ is $\mathrm{Spin}(2, 1)\simeq \SL(2, \mathbb R)$. The anti-Hermitian generators of the Lorentz group satisfy the commutation relations
\be
\label{eq:Lorentz-group-algebra-comm}
	[M^{\mu\nu},M^{\r\s}]=\eta^{\nu\r}M^{\mu\s} + \eta^{\mu \s} M^{\nu \r}-\eta^{\mu\r}M^{\nu\s} -\eta^{\nu\s}M^{\mu\r} \,,
\ee
where the Lorentzian metric signature is chosen to be	$\eta^{\mu\nu}=\eta_{\mu\nu}=\text{diag}(-1, 1, 1)$.  The spinor representations are constructed using the gamma-matrices $\g^\mu$, which satisfy the usual relations
\be
	\g^\mu \g^\nu + \g^\nu \g^\mu = 2\eta^{\mu\nu}\,.
\ee
Explicitly, we choose
\be
\label{eq:gamma-matrices-convention}
	(\gamma^0)^\a{}_\b=\begin{pmatrix}
		0 & 1 \\ -1 & 0
	\end{pmatrix},\qquad
	(\gamma^1)^\a{}_\b=\begin{pmatrix}
	0 & 1 \\ 1 & 0
	\end{pmatrix},\qquad
	(\gamma^2)^\a{}_\b=\begin{pmatrix}
	1 & 0 \\ 0 & -1
\end{pmatrix}\,.
\ee
The Lorentz generators are then represented by the matrices
\be
\label{eq:lorentz-generators}
	\p{\cM^{\mu\nu}}^\a{}_\b=\frac{1}{4}\p{[\g^\mu,\g^\nu]}^\a{}_\b\,.
\ee
Note that the representation matrices $\cM$ are real since the $\g$-matrices are. These matrices satisfy the same commutation relations as~\eqref{eq:Lorentz-group-algebra-comm} and preserve the symplectic form
\be
\label{eq:symplectic-form-convention}
	\O_{\a\b}=\O^{\a\b}=\begin{pmatrix}
			0 & 1\\
			-1 & 0
	\end{pmatrix}\,.
\ee
The elements of the spinor irrep are real two-dimensional vectors with upper indices $s^\a$. We raise and lower indices by using the symplectic form $\O$
\be
	s_\a=\O_{\a\b}s^\b,\qquad s^\a=s_\b \O^{\b\a}.
\ee
A general finite-dimensional irrep of $\SL(2,\R)$ is labeled by a non-negative (half-)integer spin $j$. The elements of these representation are symmetric tensors of the form
\be
	T^{\a_1\cdots\a_{2j}}.
\ee
All these irreps are real since the spinor irrep is real.

A local operator of spin $j$ is a tensor
\be
	\cO^{\a_1\cdots\a_{2j}}(x),
\ee
symmetric in indices $\a_i$. Its transformation properties under the Lorentz group are specified by the commutation relation
\be\label{eq:MOcommutator0}
	[M^{\mu\nu},\cO^{\a_1\cdots\a_{2j}}(x)]=(x^\nu \ptl^\mu-x^\mu \ptl^\nu)\cO^{\a_1\cdots\a_{2j}}(x)-\sum_{k=1}^{2j} (\cM^{\mu\nu})^{\a_k}{}_\b \cO^{\a_1\cdots \a_{k-1}\b\a_{k+1}\cdots \a_{2j}}(x).
\ee
In the main text we often use the index-free notation
\be
	\cO(s)=s_{\a_1}\cdots s_{\a_{2j}}\cO^{\a_1\ldots \a_{2j}}\,,
\ee
where $s$ is a real spinor variable, whose components we often denote by
\be
s_{\a} \equiv \begin{pmatrix}
	\xi \\ \bar \xi
\end{pmatrix}.
\ee
For a more detailed discussion of our conventions, refer to appendix A of~\cite{Erramilli:2019njx}.

\section{Parity for the three-point and four-point structures}
\label{app:parity-conv}

In this section we clarify the meaning of parity for the tensor structures. We define the parity $\kappa$ of a local operator by
\be
	R_\mu \cO(x,s) R_\mu^{-1}=\kappa \cO(R_\mu x,\gamma_\mu s),
\ee
where $R_\mu$ is the unitary operator representing reflection in a spatial direction $\mu$ ($x^\mu\to -x^\mu$), $\mu=1,2$, and $R_\mu x$ is the appropriately reflected $x$. For $\kappa=1$ we say that the operator is parity-even, and for $\kappa=-1$ we say that the operator is parity-odd. This definition is consistent with the usual definition of parity for tensor (integer $j$) operators. 

Motivated by this definition, for a tensor structure represented by a function $f(x_i,s_i)$ of several coordinates $x_i$ and polarizations $s_i$ we define 
\be
	(R_\mu f)(x_i,s_i)\equiv f(R_\mu x_i, \g_\mu s_i).
\ee
It follows that $R_\mu^2 f=f$ and thus all structures can be split into parity-even ($R_\mu f=f$) and parity-odd ($R_\mu f=-f$).

For operators of definite parity $\kappa$, correlation functions are expanded in terms of parity-even structures if the product of operator parities is even, and in terms of parity-odd structures if the product of operator parities is odd.

\section{Clebsch-Gordan coefficients}
\label{app:CG-coeff}

In our conventions the Clebsch-Gordan coefficients are given by the formula
\be
 	\<&j_1,m_1;j_2,m_2|j,m\> = \sqrt{\frac{(2j+1)(j+j_1-j_2)!(j-j_1+j_2)!(j_1+j_2-j)!}{(j_1+j_2+j+1)!}} \nn \\ &\times \sqrt{(j+m)!(j-m)!(j_1+m_1)! (j_1-m_1)! (j_2+m_2)! (j_2-m_2)! }\nn \\ &\times \sum_k \frac{(-1)^k}{k! (j_1+j_2-j-k)!(j_1-m_1-k)!(j_2+m_2-k)!(j-j_2+m_1+k)!(j-j_1-m_2+k)!} \,,
\ee
where the sum runs over values of $k$ for which the arguments of the factorials in the denominator are non-negative.

\bibliographystyle{JHEP}
\bibliography{refs}

\end{document}